\documentclass[aps,prd,amsmath,amssymb,11pt]{revtex4}
\usepackage{graphicx}
\usepackage{dcolumn}
\usepackage{bm}

\newcommand{\be}{\begin{equation}}
\newcommand{\ee}{\end{equation}}
\newcommand{\bea}{\begin{eqnarray}}
\newcommand{\eea}{\end{eqnarray}}
\newcommand{\nn}{\cr}
\newcommand{\hf}{\frac12}
\def\journal#1#2#3#4{{#1} {\bf #2}, #3 (#4)}
\def\eq#1{(\ref{#1})}
\def\la{\langle}
\def\ra{\rangle}

\def\Tr{{\mathrm{Tr}}}
\def\ord#1{{\cal O}\left(#1\right)}
\def\mr#1{{\mathrm{#1}}}

\def\fdd#1#2#3{\frac{\delta^2#1}{\delta#2\delta#3}}

\def\dk{\Delta k}

\begin{document}
\title{First order phase transition with functional renormalization group method}
\author{S. Nagy$^1$, J. Polonyi$^2$}
\affiliation{$^1$Department of Theoretical Physics, University of Debrecen,
P.O. Box 5, H-4010 Debrecen, Hungary}
\affiliation{$^2$Strasbourg University, CNRS-IPHC, \\23 rue du Loess, BP28 67037 Strasbourg Cedex 2, France}
\date{\today}
\begin{abstract}
The renormalization group method, more specifically the Wegner-Houghton equation, is used to find first order phase transitions in a simple scalar field theory with a polynomial potential. An improved definition of the running parameters allows us to explore the renormalization group flow down to the IR end point and to locate phase transitions. Beyond the expected first order transition further radiative correction generated first and second order transitions are found. The phase diagram is reviewed by a Monte-Carlo simulation of the lattice regulated version of the theory but the serious slow down of the convergence prevents us to obtain conclusive results from the simulation.
\end{abstract}
\maketitle

\section{Introduction}
Phase transitions are defined as singularities in the relation between the infrared (IR) and the ultraviolet (UV) quantities of a system. The origin of the singularities is a subtle question because systems with finite number of degrees of freedom possess analytic scale dependence therefore display no phase transition. Phase transitions take place only in strictly infinite systems which are beyond our experimental and analytical possibilities hence they have to be considered as useful approximations for systems with a large but finite number of degrees of freedom.

The theoretical challenge of phase transitions is to find the origin of a singularity in the relation between the IR and the UV quantities. Field theories can be rendered finite by introducing an UV and an IR cutoff and phase transitions appear when one of these cutoffs is removed. Since the physical laws are known up to a finite spatial resolution only the IR cutoff can be removed in a realistic theory.

The best method to deal with such a problem is the renormalization group, devised just to follow the scale dependence. The renormalization group was used originally to send the UV cutoff to infinity in relativistic quantum field theory by keeping the physics fixed at a finite scale \cite{stuckelberg,bogo}. The method was recovered independently later in statistical mechanics to explain the emergence of power like singularities and universality at the critical points of second order phase transitions by constructing block spin variables by the decrease of the UV cutoff in finite steps, \cite{essam,widom,kadanoff}. The relation between the two procedures was understood soon \cite{wilson} and can be summarised in a simple cases of a relativistic massive quantum field theory and a spin model as follows: The models have two important scales, one is the UV cutoff, say the highest momentum $\Lambda$ and the lattice spacing $a$ in quantum field theory and lattice spin models, respectively and the other is an intrinsic IR scale, say the Compton wavelength $\lambda_C$ and the correlation length $\xi$ of the spin system. The Compton wavelength serves as a correlation length for the quantum field because a particle can not be localized within its Compton wavelength.

The intrinsic scale separates the UV and the IR scaling regimes, defined by the cutoff dependence of physical quantities in such a manner that the cutoff-dependence is weak in the IR scaling regime. This is easiest to see in the spin system where the evolution of the parameters of the Hamiltonian during an increase of the lattice spacing is due to the fluctuations within a block with length scale between the old and the new lattice spacing. Since these parameters are usually defined as the value of some special connected Green functions, one-particle irreducible vertex functions, their evolution slows down as we enter into the IR scaling regime beyond the correlation length.

We can now return to the question of the singularities. It was found that the contributions to the relation between the UV and IR quantities pile up in an approximately scale invariant manner in the UV scaling regime. Hence the diverging length of this scaling regime, $\Lambda\lambda_C=\xi/a\to\infty$, serves as the source of the singularities of quantum field theories and critical systems.

We have surveyed the origin of the singularities in continuous phase transition to motivate the question leading to the topic of this work: Where do the singularities of a first order phase transition with finite the correlation length come from? The pictured suggested here is that the continuous and the discontinuous phase transitions agree that the singularities emerge from the piling up fluctuations in a long scale window rather than from the dynamics at a given scale but differ in the scale window. The critical behaviour comes from the UV asymptotic scaling regime. The dynamics is approximatively scale invariant here because the only scale parameter here, the cutoff, is hidden by the running of the parameters of the action. The first order  phase transitions are driven by processes at a finite scale, the formation of critical droplets \cite{langer,fischer,voloshi,binder} but the singularities appear only after a long non scale-invariant evolution in the IR scaling regime, in the thermodynamical limit.

The goal of this work is to find a first order transition in a simple $\phi^6$ Euclidean scalar model in three dimensions by the help of the renormalization group method. This method is usually employed locally, around an UV fixed point, to remove the UV cutoff from quantum field theories and establish universal critical exponents. Another fixed points such as the Wilson-Fisher \cite{wilsonf} and the Kosterlitz-Thouless \cite{kosterlitz} have been found useful to understand the origin of separatrices in the renormalization group flow at phase boundaries. A more systematic global extension of the renormalization group flow, the global renormalization group \cite{grg}, shows the richness of the scale dependence in more realistic theories with several intrinsic scales by offering different classification schemes for the same observable algebra. The proposed view of first order transitions is motivated by this approach where the emphasis lies on the scale dependence beyond a simple asymptotic scaling analysis close to a fixed point.

The renormalization group method becomes more powerful by exploiting a new small parameter, the step size of the scale change. When the gliding UV cutoff, denoted below by $k$, is defined in momentum space then its change $\dk$ can be as small as $\ord{1/L}$ in units of $c=\hbar=1$ where $L$ is the IR cutoff, the size of the system. The beta functions, the derivative of the the running parameters with respect to $k$, are given by the one-loop expression in the infinitesimal blocking limit $\dk\to0$ because the higher loop contributions are $\ord{\dk}$ \cite{wilson,wh,polchinski}. A slightly different way to use such a possibility of arriving at exact evolution equations is to give up the plan of keeping the physics independent of the running cutoff and to follow the evolution of the effective action along an artificial trajectory in the space of theories connecting a soluble theory with the physical one by the help of an IR cutoff \cite{wetterich}.

There is an arbitrary step in setting up the renormalization group method. The point is that the running parameters of the action are not uniquely defined due to a conflict: On the one hand, the analytic form of the action must be fixed for any practical calculation, and the other hand, the elimination of degrees of freedom during the lowering of the UV cutoff induces infinitely many new terms to the action. In other words, while we can handle actions with a limited number of parameters the renormalization group keeps generating infinitely many new one. Thus the definition of the new parameters is an overdetermined problem. This is actually the strength rather than a drawback of the renormalization group method because it forces us to find an appropriate definition of few running parameters which characterize a large number of physical processes. This step, the choice of the blocking procedure in Kadanoff's scheme or the subtraction procedure in the perturbative renormalization in quantum field theory, is not unique and needs optimisation.

There have already been a number of works about the use of the renormalization group method for first order phase transitions. The indication that this method can capture discontinuous phase transitions came from the finding that the radiative corrections of a gauge field generate a first order transition in superconductors  \cite{halpernl} and in scalar QED \cite{coleman}. In the same time a first order transition was found in simpler models, in anisotropic cubic systems \cite{wallace} and its relation to the loss of a fixed point was noted in ref. \cite{rudnick}. A sufficient condition of a first order transition was identified as a fixed point with a strongly relevant operator \cite{nienhuis}. The dynamical renormalization group was used, as well, to describe first order transitions by non-critical fixed point \cite{zhong}. The essential singularities of the coexistence region can be recovered as the renormalized trajectory passes a fixed point \cite{klein}. The rounding effect of the finite size scaling on first order transition was the subject of ref. \cite{privman}. The first order phase transition may lead to multi-valued thermodynamical potential in certain approximations. In a similar manner the blocking transformation may become singular in the vicinity of first order transition \cite{hasenfratz}. This result underlines the importance of the proper choice of the subtraction procedure, our main concern here. The temperature driven first order transitions can be classified by the help of the renormalization group \cite{liang}. The first order transitions were described in scalar models by following the IR cutoff-dependence of the effective action in refs. \cite{tetradisw,teradis}. A part of the phase structure of the $\phi^6$ scalar model discussed here has been found by the help of the traditional subtraction procedure, based at vanishing field \cite{jakubszyk}. Another application of the functional renormalization group method covers first order transitions in lattice-gas models \cite{parola}. The first order transition occuring in fermionic system has been discussed \cite{gersch}, too.

The appropriate choice of the subtraction point holds the key to recover first order phase transitions in the IR scaling regime and our choice is presented in section \ref{solvings}. As a simple consistency check of the new subtraction procedure the recovery of the Wilson-Fisher fixed point is shown in section \ref{wffpf}. The renormalization group flow of the new subtraction scheme generates a complicated phase structure with several radiation correction induced first and second order transitions, these results are reported in section \ref{firsttrs}. Such a rich structure is unexpected and a Monte-Carlo simulation was carried out to clarify the situation. As discussed in section \ref{lattices} the simulation slows down enormously just in the interesting regions of the parameter space and it can neither prove nor disprove the predictions of the renormalization group method. Finally our results are summarized in section \ref{summarys}.

\section{Solving a theory by the renormalization group}\label{solvings}
The renormalization group method is used below to find the evolution of the bare action during the lowering of the UV momentum space cutoff $k$ of an Euclidean scalar field theory given by the action
\be\label{baction}
S[\phi]=\int d^dx\left[\hf(\nabla\phi(x))^2+U(\phi(x))\right]
\ee
where the potential is a $N_U$-th order polynomial,
\be\label{bpot}
U(\phi)=\sum_{n=1}^{N_U}\frac{g_n}{n!}\phi^n.
\ee
The only dimension of the model is expressed in units of the initial cutoff by using the bare theory $k_i=1$ and $g_n=g_{Bn}$ as initial condition for the lowering of $k$.

\subsection{Gliding UV cutoff}
The blocking in momentum space, the lowering $k\to k-\dk$ of the UV cutoff, is supposed to preserve the  partition function
\be\label{partfnct}
Z=\int D[\phi]e^{-S_k[\phi]}
\ee
hence $S_{k-\dk}$ is found by integrating out the modes with wave vector $k-\dk<|p|<k$ \cite{wh},
\be\label{blocking}
e^{-S_{k-\dk}[\phi]}=\int D[\phi']e^{-S_k[\phi+\phi']}
\ee
where the field variables $\phi(x)$ and $\phi'(x)$ are non-vanishing for $|p|<k-\dk$ and $k-\dk<|p|<k$ in the momentum space, respectively. The evolution equation
\be
S_k[\phi]-S_{k-\dk}[\phi]=-\hf\Tr\ln\left[\fdd{S_k[\phi]}{\phi'}{\phi'}\right]+\ord{\dk}
\ee
obtained by ignoring saddle-points contains corrections beyond the simple one-loop expression because the loop integration is over a shell of $\dk$ thickness in momentum space. These corrections are vanishing in the small step size limit governed by the functional differential equation
\be\label{evoleqfdk}
\partial_kS_k[\phi]=-\lim_{\dk\to 0}\frac1{2\dk}\Tr\ln\left[\fdd{S_k[\phi]}{\phi'}{\phi'}\right].
\ee
The right hand side is finite since the functional trace is taken over a momentum shell of thickness $\dk$.

To arrive at a manageable problem one employs the only approximation of the calculation, the projection of the evolution equation onto a restricted functional ansatz space for the blocked bare action. Such a functional space is usually defined by the help of the Landau-Ginzburg double expansion in Euclidean space-time, by assuming that the field has long distance fluctuations compared to the cutoff with small amplitude. To keep our calculation as simple as possible we consider a single component scalar field theory within the local potential approximation where the bare action is assumed to be of the form \eq{baction}-\eq{bpot}. The value $N_U=6$ will be used in the numerical work with $d=3$ and the potential is assumed to be symmetric, $g_{2n+1}=0$, except if it is stated contrary.

\subsection{IR effective theory}\label{ireffths}
The promise of eq. \eq{evoleqfdk} is to deliver a more dilute theory with a lower cutoff which is easier to handle. To find a suitable approximation for this reduced theory we introduce an IR cutoff by placing the IR effective theory into a box of size $L=2N_p\pi/k$, $N_p$ being an integer cutoff parameter, equipped with periodic boundary conditions. One has $N^d_p$ modes in the path integral \eq{partfnct} for the field variable
\be
\phi(x)=\sum_{m_\mu=0}^{N_p-1}\phi_ne^{ip_mx}
\ee
where $p^\mu_m=2\pi m^\mu/L$, with the action density
\be
\frac{S[\phi]}{L^d}=-\hf\prod_{j=1}^2\sum_{m_{j,\mu}=0}^{N_p-1}\delta_{m_1+m_2,0}p_{m_1}p_{m_2}\phi_{m_1}\phi_{m_2}+\sum_n\frac{g_n}{n!}\prod_{j=1}^n\sum_{m_{j,\mu}=0}^{N_p-1}\delta_{\sum_jm_j,0}\phi_{p_{m_1}}\cdots\phi_{p_{m_n}}.
\ee

By assuming that the cutoff $k$ is low enough and no IR singularities left we take a radical step and restricts the size to $L=2\pi/k$ leaving a single mode $\phi_0$ without kinetic energy,
\be\label{zappr}
Z=\int d\phi_0e^{-(\frac{2\pi}{k})^dU_k(\phi_0)}.
\ee
This approximation is easier to understand in Kadanoff's real space blocking \cite{kadanoff} where the field variable is defined on a space-time lattice and one increases the lattice spacing by a fixed ratio $a\to sa$. The blocked action, defined for the more dilute blocked lattice, is given by
\be
e^{S_{as}[\phi']}=\prod_n\int d\phi_ne^{-S_a[\phi]}\prod_{n'}\delta(\phi'_{n'}-B_{n'}[\phi])
\ee
where $B_{n'}[\phi]$ denotes the blocked field variable of the block at $n'$. The approximation \eq{zappr} correspond to have a single site blocked lattice. One looses completely the resolution in the space-time but retains enough information in the field dynamics. This approximation motivates our next step, the choice of the blocking procedure.

\subsection{Subtraction procedure}
The traditional subtraction scheme of the local potential ansatz with an arbitrary potential $U(\phi)$ is to evaluate the evolution equation \eq{evoleqfdk} at the most IR field configuration, $\phi(x)=\phi$. The result is the Wegner-Houghton equation \cite{wh},
\be\label{wh}
\dot U(\phi)=-\frac{\Omega_d}{2(2\pi)^d}k^d\ln(k^2+U^{(2)}(\phi)),
\ee
where $\Omega_d$ stands for the $d$-dimensional solid angle, the dot denotes the derivative with respect to  $t=\ln k/k_i$, $k_i$ stands for the initial value of the cutoff and $U^{(n)}(\phi)=\partial^n_\phi U(\phi)$. One may use different way to extract the evolution of the potential from eq. \eq{evoleqfdk} like there are different definitions of the block spin in real space blocking. A choice is better if the resulting parameters cover better the physics at the scale of the cutoff.

We note at this point that the Wegner-Houghton equation can be rewritten for a transformed field variable $\rho=f(\phi)$ in an obvious manner. But the resulting evolution equation can not be interpreted in terms of the composite operator $\rho$ since $\la f(\phi)\ra\ne f(\la\phi\ra)$ for non-linear transformations. Nevertheless such a transformation might be useful if it belongs to a symmetry of the potential, $U(\phi)=U(f(\phi))$ because the beta functions remain invariant and one can separate the Higgs and the Goldstone modes. However the singularities of the parametrization exclude some values, for instance a spontaneous symmetry breaking cannot be established by $\rho=\phi^2$.

The order parameter is the location of the absolute minimum of the potential $\phi_a$ at the IR end point, $k=0$. Hence the solution of the Wegner-Houghton equation places the phase transition at a separatrix among the renormalized trajectories, where an infinitesimal change of the initial conditions at $k_i$ induces finite changes at $k=0$. Such a characterization of the phase transition is possible only in the thermodynamical limit since for a large but finite quantization box $L<\infty$ the momentum spectrum is discrete, the minimal step size at the blocking \eq{blocking} is $\dk=\ord{1/L}$, and there are corrections to the Wegner-Houghton equation.

\subsection{Finite $N_U$}
The restriction of the potential to a $N_U$-th order polynomial makes the choice of the subtraction procedure more important because the dynamics is now characterized by fewer parameters. The first choice would be to evaluate the Wegner-Houghton equation at $N_U$ values of the field. But it is not obvious how to pile up these points in the regions dominated by the quantum fluctuations. A better choice is to expand the potential at a base point $\phi_b$
\be
U(\phi)=\sum_{n=1}^{N_U}\frac{G_n(\phi_b)}{n!}(\phi-\phi_b)^n
\ee
and use the first $N_U$ derivatives of the Wegner-Houghton equation to define the beta function of $G_n(\phi_b)$ by the equation
\be\label{betafncts}
\beta_n(\phi_b)=\partial^n_\phi\dot U(\phi_b).
\ee

The renormalized trajectories depend on the choice of $\phi_b$ owing to the non-polynomial nature of the right hand side of the evolution equation in $U^{(2)}$. To find the best choice of the base point one needs information about the way the parameters are used later, below the current cutoff. The simplest tree-level approximation within the blocked theory is the choice $N_p=1$, described above in section \ref{ireffths}, where the beta functions are calculated at the absolute minimum $\phi_a$ of the potential. The Wegner-Houghton equation taken at the evolving minimum,
\be\label{whm}
\dot U(\phi_a)=\dot\phi_aU^{(1)}(\phi_a)-\frac{\Omega_d}{2(2\pi)^d}k^d\ln[k^2+U^{(2)}(\phi_a)],
\ee
yields the evolution equation
\be\label{phim}
\dot\phi_a=-\frac{\beta_1}{G_2}
\ee
for the minimum and
\be\label{gevm}
\dot G_n=\beta_n-\beta_1\frac{G_{n+1}}{G_2}
\ee
for the parameters of the potential. The parameters corresponding to the vanishing base point will be used frequently and are denoted by $g_n=G_n(0)$.

The guidance of the tree-level approximation in the thinned blocked theory leads to serious difficulties as soon as $\phi_a$ changes discontinuously during the evolution. In fact, a necessary condition of the local uniqueness of the solution of the differential equation $\dot x=f(x,t)$ is the continuity of the right hand side in $x$. Hence the renormalized trajectory, the integral of the evolution equation, is not unique when $\phi_a$ jumps. A simple numerical manifestation of this problem is the ill-defined nature of the beta functions when the absolute minimum becomes degenerate. One would have thought that this case can safely be ignored for the blocking steps $k\to k-\dk$ with finite $\dk$ but such a naive argument proves to be wrong, cf. Fig. \ref{typtrajf}(b) below.

A discontinuous jump of the beta functions defined at $\phi_a$ is unacceptable on physical grounds, too, and more involved physical processes to be taken into account by an extension of the ansatz for the blocked action or the subtraction procedure should smear out these discontinuity. In fact, it is a basic tenet of the renormalization group method that the dynamics of any finite scale window is regular and the possible singularities of phase transitions build up from the diverging size of the scale window.

The discontinuity generated by the tree-level approximation of the thinned blocked theory can be smeared out by using the average beta functions
\be\label{avbf}
\beta_n(\phi_b)=\frac{\int D[\phi]\beta_n(\phi_b,\phi_0)e^{-S_k[\phi]}}{\int D[\phi]e^{-S_k[\phi]}},
\ee
where $\phi_0$ denotes the homogeneous component of $\phi(x)$ and $\beta_n(\phi_b,\phi_0)$ stands for the beta function obtained by expanding the Wegner-Houghton equation at $\phi_0$ and transforming its beta function for the Taylor expansion parameters $G(\phi_b)$ of the potential defined to the base point $\phi_b$. This average simplifies to
\be
\beta_n(\phi_b)=\frac{\int d\phi_0\beta_n(\phi_b,\phi_0)e^{-(\frac{2\pi}{k})^dU_k(\phi_0)}}{\int d\phi_0e^{-(\frac{2\pi}{k})^dU_k(\phi_0)}}
\ee
by the help of the approximation \eq{zappr}. A further simplification can be made by applying the saddle-point approximation,
\be\label{weightavb}
\beta_n(\phi_b)=\frac{\sum_j\beta_n(\phi_b,\phi_j)b_j}{\sum_jb_j}
\ee
where
\be
b_j=\frac{e^{-c_s(\frac{2\pi}{k})^dU_k(\phi_j)}}{\sqrt{U^{(2)}_k(\phi_j)}},
\ee
and the points $\phi_j$ are local minima of the potential, $U^{(1)}_k(\phi_j)=0$, $U^{(2)}_k(\phi_j)>0$. The   dimensionless parameter of the subtraction procedure $c_s$ is introduced to diagnose the sensitivity of the renormalized trajectory on the choice of the absolute minimum but its value is kept at $c_e=1$ in the results presented below.

The transformation of the beta function to a different base point $\phi_b\to\phi'_b$ is made by the linear transformation of the Taylor coefficients,
\be
G_m(\phi'_b)=\sum_{n=1}^{2N_U}A_{m,n}(\phi'_b-\phi_b)G_n(\phi_b)
\ee
with
\be
A_{m,n}(\phi)=\begin{cases}\frac{\phi^{n-m}}{(n-m)!}&m\le n\cr0&n<m\end{cases}.
\ee
This results from a direct and trivial calculation or from the introduction of the ladder matrix $S_{mn}=\delta_{n+1,m}$ with the properties $S^j_{mn}=\delta_{n+j,m}$ and $S^{N_U}=0$ allowing us to write $A(\Phi)=e^{\Phi S}$. The average beta function at $\phi_b=0$
\be\label{expmix}
\beta_n(0)=-\frac{\alpha_d}2k^d\frac{\sum_jb_j\sum_mA_{nm}(-\phi_j)\partial_\phi^m\ln[k^2+U^{(2)}(\phi_j)]}{\sum_jb_j}
\ee
is used to integrate the evolution equation $\dot g_n=\beta_n(0)$.

Few remarks are in order at this point:

1. The use of the minima assures $G_2\ge0$ and excludes the instability $k^2+G_2\le0$ during the evolution.

2. The summation in \eq{expmix} over the different minima is needed since the approximation \eq{zappr} keeps the volume finite and the vacuum unique for $k>0$.

3. The weighted average of the beta functions at different minima is more and more dominated by the absolute minimum as the the IR end point $k=0$ is approached. Phase transition occurs when the minima are degenerate in the limit $k\to0$.

4. In case of symmetric initial condition the symmetry is preserved during the evolution. Nevertheless it is advised to reinforce the symmetry with respect to $\phi\to-\phi$ and to cancel the beta functions of odd orders $g_{2n+1}$ because the small rounding errors at small $n$ appear as relevant and their growth in the UV scaling regime lead to the gradual loss of the symmetry.

5. While the solution of the Wegner-Houghton equation \eq{wh} for a transformed field variable $\rho=f(\phi)$ can be obtained by the help of a trivial rewriting of the differential equation for $\rho$ ambiguities appear for finite $N_U$. In fact, the result of the projection of the right hand side of the equation depends on the choice of the base point $\phi_b$. For instance the transformation $\rho=\phi^2$ is advantageous for symmetric potential $U(\phi)=V(\rho)$ whose evolution equation is
\be
\dot V=-\frac{\alpha_d}2k^d\ln(k^2+2V^{(1)}+4\rho V^{(2)}).
\ee
However the truncation of the potential $V(\rho)$ to a polynomial of order $N_U/2$,
\be
U(\phi)=V(\rho)=\sum_{n=1}^{N_U/2}\frac{r_n}{n!}\rho^2,
\ee
yields the beta functions
\be
\dot r_n=-\frac{\alpha_d}2k^d\partial_\rho^n\ln(k^2+2V^{(1)}+4\rho V^{(2)})
\ee
and this set of equation is not equivalent with the integration of the evolution equation for the first $N_U$ coefficients $G_n$ with the beta functions \eq{betafncts}. This can easily be seen by noting that we have the first $N_U/2$ and the $N_U$ derivatives of the logarithm function in the $\rho$ and the $\phi$ parametrization, respectively. Another, more illuminating difference between the two parametrization is visible by comparing the beta functions with their perturbation expression: The usual Feynman graphs are recovered only for the $\phi$ parametrization because there is no Wick-theorem for the composite operator $\rho$.

\section{Fixed points}\label{wffpf}
Important features of the phase structure follow from the fixed points of the evolution equation. The fixed  potential satisfies the dimensionless Wegner-Houghton equation,
\be\label{fpwh}
d\tilde V-d_\phi\tilde\phi\tilde V^{(1)}=-\frac{\alpha_d}2\ln(1+\tilde V^{(2)}),
\ee
where $d_\phi=d/2-1$ denotes the mass dimension of the field and the tilde indicates that the dimension is removed by the help of the cutoff, $\phi=k^{d_\phi}\tilde\phi$, $g_n=k^{d-nd_\phi}\tilde g_n$ and $U(\phi)=k^d\tilde V(\tilde\phi)$. The solutions of the fixed point equations form a two-dimensional manifold. Apart of two discrete points, the trivial Gaussian and the non-trivial Wilson-Fisher fixed points, these fixed-potentials correspond to strongly non-perturbative theories with diverging repulsion at finite particle density \cite{morris1,halpern,morris2}. As long as our intuitive ideas about quantum field theories are taken from the perturbation expansion and its partial resummation we have to be contended by the two discrete fixed points.

The approximation of the potential by a finite order polynomial leads to the evolution equation $\dot{\tilde G}_n=\tilde\beta_n$ for the dimensionless parameters of the potential at a fixed expansion point $\tilde\phi_b$ and results the fixed point equation
\be\label{fixpb}
(d-nd_\phi)\tilde G_n=\tilde\beta_n.
\ee
A further simplification at the fixed point where $k$ can be arbitrarily small is that the exponential weighted beta function \eq{expmix} can be replaced by the beta function calculated at the absolute minimum.

\begin{figure}
\includegraphics[width=5cm,angle=-90]{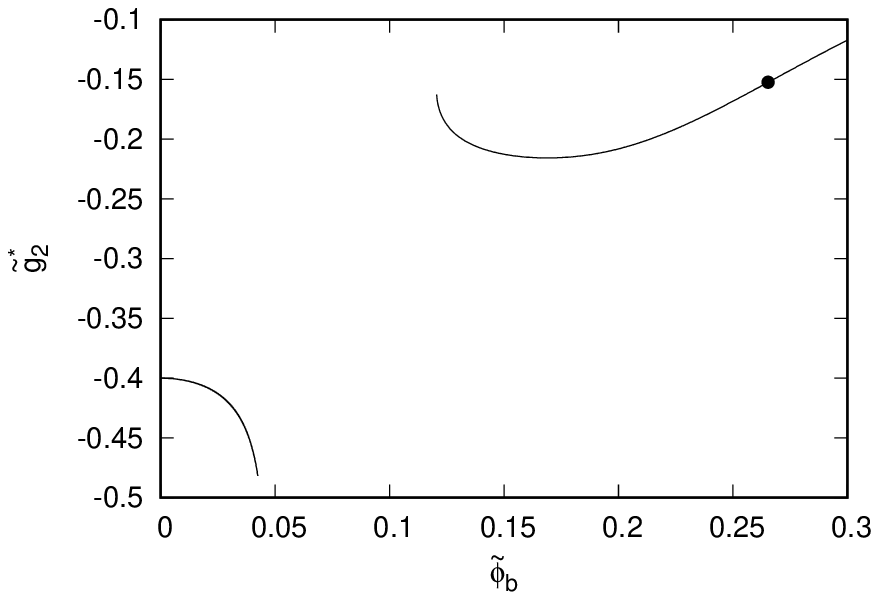}\hskip1cm
\includegraphics[width=5cm,angle=-90]{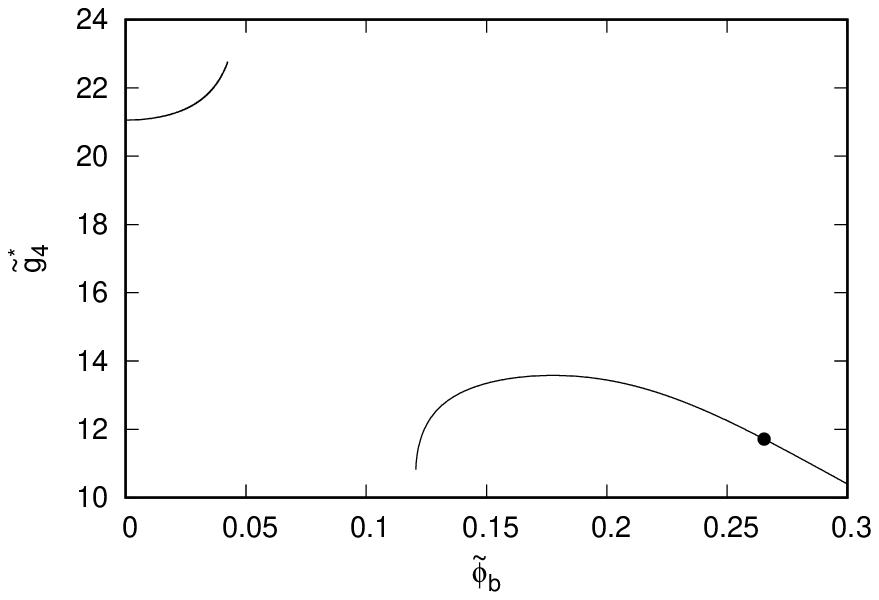}

\hskip.2cm(a)\hskip8cm(b)

\includegraphics[width=5cm,angle=-90]{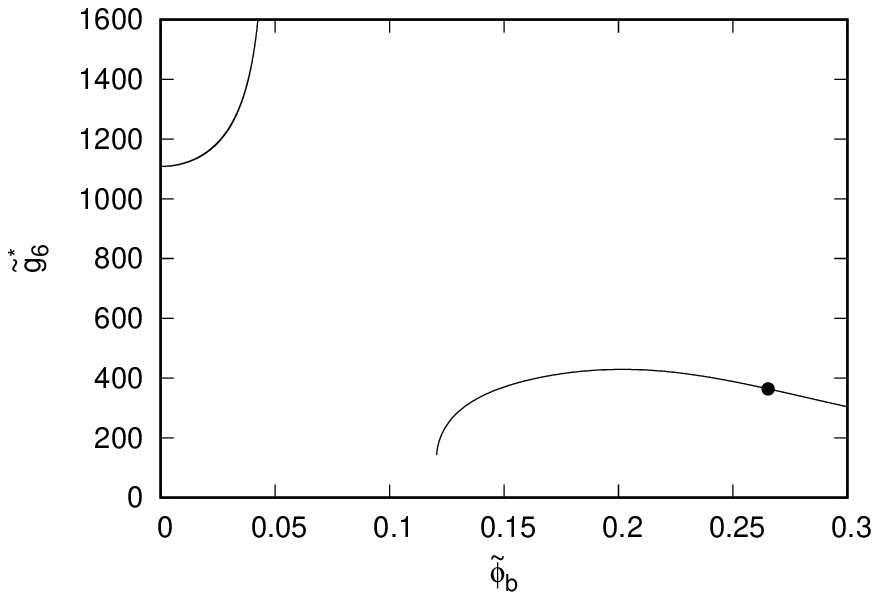}\hskip1cm
\includegraphics[width=5cm,angle=-90]{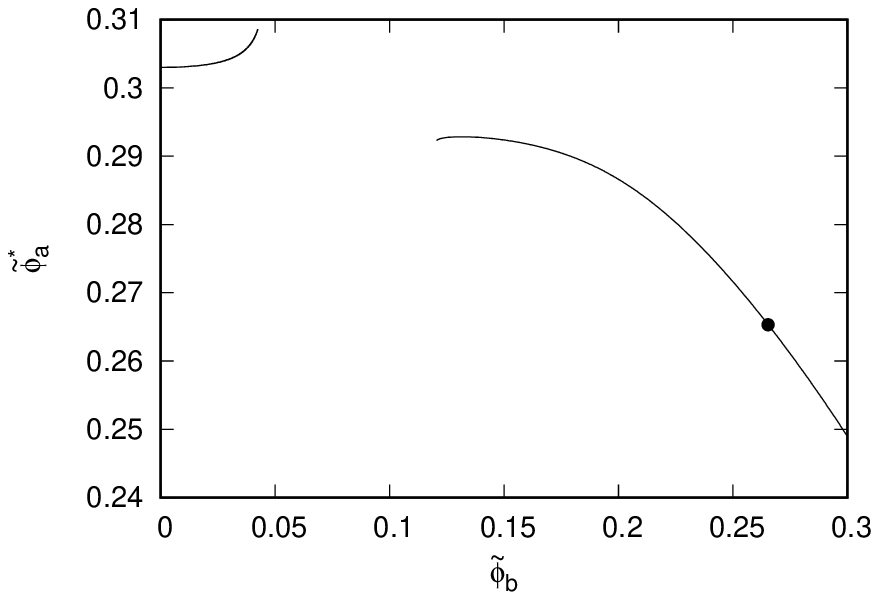}

\hskip.2cm(c)\hskip8cm(d)
\caption{The parameters of the sixth order symmetric polynomial, (a): $\tilde g^*_2$, (a): $\tilde g^*_4$, (a): $\tilde g^*_6$, and (d): the minimum $\tilde\phi_a^*$ as the function of the base point of the expansion $\tilde\phi_b$.}\label{fixedpf}
\end{figure}

One finds easily the trivial Gaussian fixed point at $\tilde\phi_b=0$. The values of $\tilde g^*_2$, $\tilde g^*_4$ and $\tilde g^*_6$ at the non-trivial fixed point for symmetric potential, $g^*_1=g^*_3=g^*_5=0$, together with their minimum are shown in Fig. \ref{fixedpf} as the function of the base point $\tilde\phi_b$. One finds the traditional Wilson-Fisher fixed point for $\tilde\phi_b=0$. The corresponding potential has a non-trivial minimum hence this fixed point is inconsistent, its potential is obtained by expanding around an unstable base point. The extension of the fixed point, the curve leaving the Wilson-Fisher fixed point, exists for small $\tilde\phi_b$ and the potential becomes complex for $\tilde\phi_b>0.04$, far from the actual minimum.

One can find a consistent fixed point by choosing the base point of the expansion at the minimum of the potential. The solution of eq. \eq{fixpb} together with this auxiliary condition, shown by the dot in Fig. \ref{fixedpf}, can be extended for a larger interval of the base points but this manifold of solutions is substantially different than the Wilson-Fisher fixed point manifold. Hence the renormalized trajectories of the present subtraction procedure passing in the vicinity of the non-trivial fixed point are qualitatively different than the traditional one, guided by the beta functions calculated around the original Wilson-Fisher fixed point.

The fixed point Wegner-Houghton equation \eq{fpwh} possesses asymmetric solutions, $\tilde V(\tilde\phi)\ne\tilde V(-\tilde\phi)$. However no non-trivial sixth order fixed point polynomial was found by relaxing the symmetry of the potential indicating that the critical phenomenon is strictly related to spontaneous symmetry breaking in this model.

\section{Phase structure}\label{firsttrs}
The model supports a number of phase transitions with and without spontaneous symmetry breaking with symmetric $U(\phi)=U(-\phi)$ and with full potential $U(\phi)\ne U(-\phi)$, respectively. As of the former, there is the traditional second order phase transition with bare parameters with initial conditions $g_{B2}<0$, $g_{B4},g_{B6}\ge0$. The absolute minimum of the potential depends in a continuous manner in the bare parameters around this transition hence one expects only quantitative changes compared to the traditional subtraction procedure at $\phi_b=0$ around this transition. Therefore we devote our attention to other possible phase transitions.

\subsection{Tree-level phase diagram}\label{freephds}
\begin{figure}
\includegraphics[width=6cm,angle=-90]{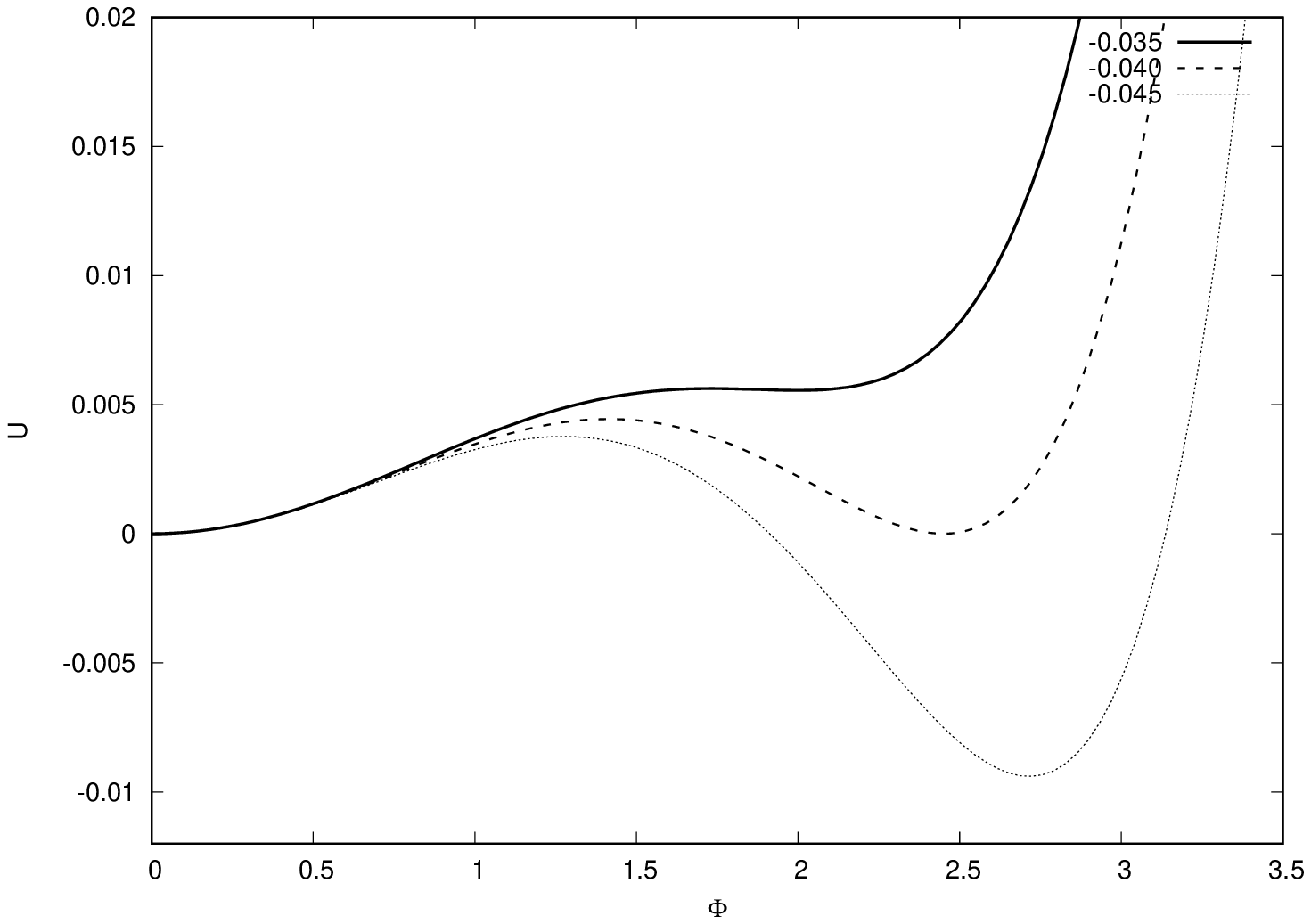}
\caption{The symmetric potential with $g_2=0.01$, $g_4=-0.035,-0.04,-0.045$, and $g_6=0.1$. The minima are degenerate for $g_4=g_{4tr}=-0.04$.}\label{potf}
\end{figure}

The simple rule, namely that a the term $\phi^n$ in the potential with larger $n$ becomes dominant for larger values of $|\phi|$, predicts a number of tree-level phase transitions. Let us consider first a symmetric potential $U(\phi)=U(-\phi)$ where the quadratic term is dominant around zero hence the point $g_2=0$ is a second order transition. The higher power terms are important for larger field hence one may have first order transition if two non-Gaussian monomials compete. The stability of the model requires that the highest power come with positive coefficient hence we need at least a sixth order potential,
\be
U(\phi)=\frac{g_2}2\phi^2+\frac{g_4}{4!}\phi^4+\frac{g_6}{6!}\phi^6,
\ee
with $g_2,g_6>0$ and $g_4<0$ for a first order transition. The non-trivial minima appear for $g_4<-\sqrt{6g_2g_6/5}$ with absolute magnitude
\be\label{locmin}
\phi_m=\sqrt{10\frac{-g_4+\sqrt{g^2_4-\frac65g_2g_6}}{g_6}}
\ee
which are the absolute minima as long as $g_4<-\sqrt{8g_2g_6/5}$. The potential at the first order transition satisfies $g_4=-\sqrt{8g_2g_6/5}$ with $G_2(\phi_m)=4g_2$, the mass is doubled as we cross the transition line from the symmetric to the symmetry broken phase. The three dimensional parameter space $(g_2,g_4,g_6)$ with $g_6\ge0$ is cut into two parts by the second and the first order transition surfaces, $g_{2tr}=0$ and $g_2>0$, $g_{4tr}=-\sqrt{8g_2g_6/5}$, respectively. The potential is plotted on Fig. \ref{potf} for three typical set of parameters.

The appearance of a condensate in a theory with symmetric potential amounts to a spontaneous symmetry breaking. But phase transitions may occur with asymmetric potential without spontaneous symmetry breaking, too. It is easy to see for instance that one can fine tune a quartic asymmetric potential either to a critical point where the curvature at the minimum tends to zero or to a first order transition where $\phi_a$ jumps. The phases with and without symmetry breaking differ in the vertex structure of the fluctuations around $\phi_a$. However the difference of the two phases of a theory with asymmetric potential is only quantitative since $\phi_a\ne0$ in both cases.

The tree-level phase structure is not reliable in three dimensions since the fluctuations strongly deform the transition surface. Furthermore they generate latent heat where the order parameter is discontinuous and may produce multi-critical points. The modification of the phase structure by fluctuations is referred to as radiative corrections induced phase transitions. An additional complications of the first order transitions is that they support tree-level fluctuations, the droplet dynamics, a highly non-trivial saddle-point structure. It is an interesting question whether the renormalization group evolution equation which were obtained by ignoring the possible saddle-point contributions can give account of the coexistence region, dominated by the  droplets.

\subsection{Beta functions}
The beta functions \eq{betafncts} assume a particularly simple form for a symmetric potential at $\phi_b=0$,
\bea\label{sbeta}
\beta_2(0)&=&-\frac{\alpha_d}2k^d\frac{g_4}{k^2+g_2},\nn
\beta_4(0)&=&-\frac{\alpha_d}2k^d\frac{g_6(k^2+g_2)-3g_4^2}{(k^2+g_2)^2},\nn
\beta_6(0)&=&-\frac{\alpha_d}2k^d\frac{15g_4[2g_4^2-g_6(k^2+g_2)]}{(k^2+g_2)^3},\nn
\beta_1(0)&=&\beta_3(0)=\beta_5(0)=0
\eea
allowing a qualitative understanding of the renormalized trajectories. However these beta function can not be used for strong fluctuations because the large value of the propagator around the trivial vacuum $1/(k^2+g_2)$ drives $g_6$ to negative values indicating a serous inconsistency of the subtraction procedure based at $\phi_b=0$.

The instability is avoided by expanding the Wegner-Houghton equation around the minimum of the potential however this comes with a high price, the resulting beta functions are much more complicated. The expression of $\beta_n$ contains the first $n$ derivative of the right hand side of the Wegner-Houghton equation \eq{wh}. By bringing these contributions to a common denominator one generates rather complicated expressions involving fourth order polynomials of $g_2$ with coefficients up to approximately $10^9$. The inspection of the  analytical structure confirms the triviality of the IR scaling laws below $k\approx\sqrt{G_2(\phi_a)}$ and the highly non-trivial dependence on the other parameters $G_n(\phi_a)$ suggests the existence of several further characteristic scales. The global feature of the renormalized trajectory is rendered even more complicated by the possibility that a jump of $\phi_a$ may place the system at the other side of such an intrinsic scale. It is not practical to reproduce the expression of the beta functions here and we content ourself with showing it for the sake of an example at the tree-level first order transition, $g_4=g_{4tr}$, calculated at the non-trivial minimum \eq{locmin},
\bea\label{cbeta}
\beta_1(\phi_m)&=&-\alpha_dk^d3\sqrt3\left(\frac25\right)^\frac14\frac{g_2^\frac34g_6^\frac14}{k^2+4g_2},\\
\beta_2(\phi_m)&=&-\alpha_dk^d\frac1{\sqrt{10}}(-13k^2+56g_2)\frac{\sqrt{g_2g_6}}{(k^2+4g_2)^2},\nn
\beta_3(\phi_m)&=&-\alpha_dk^d\frac{\sqrt3}{2^\frac145^\frac34}(5k^4-194k^2g_2+440g_2^2)\frac{g_2^\frac14g_6^{-\frac14}}{(k^2+4g_2)^3},\nn
\beta_4(\phi_m)&=&-\alpha_dk^d\frac1{10}(5k^6-1674k^4g_2+20064k^2g_2^2-32608g_2^3)\frac{g_6}{(k^2+4g_2)^3},\nn
\beta_5(\phi_m)&=&-\alpha_dk^d\frac{\sqrt3}5\left(\frac25\right)^\frac14(-725k^6+32520k^4g_2-210480k^2g_2^2+270976g_2^3)\frac{g_2^\frac34g_6^\frac54}{(k^2+4g_2)^5},\nn
\beta_6(\phi_m)&=&-\alpha_dk^d\frac3{\sqrt{10}}(-165k^8-25948k^6g_2-485376k^4g_2^2+2107584k^2g_2^3-2256128g_2^4)\frac{\sqrt{g_2}g_6^\frac32}{(k^2+4g_2)^3}.\nonumber
\eea
The final averaged beta function
\be
\beta_n(0)=\sum_mA_{nm}(-\phi_m)\beta_m(\phi_m)
\ee
can generate highly complex renormalized trajectories.

The UV and the IR scaling laws correspond to the cutoff interval where $\tilde G_2=G_2/k^2<1$ and $\tilde G_2>1$, respectively. Hence $\tilde G_2$ can be considered as the measure of the IR-ness of the scaling laws. It usually increases monotonously during the evolution and the scale interval $0<k<k_i$ is split into two parts separated by the UV-IR crossover at $\tilde G_2=1$. Another equivalent definition of the the cutoff at the UV-IR crossover is $k_{cross}=m$ where $m=\sqrt{G_2}$ is interpreted as the mass of the elementary excitations. The asymptotic UV and IR scaling simplifies by replacement $1/(k^2+G_2)$ by $1/k^2$ and $1/G_2$ in the beta functions, respectively. The evolution is slow in the IR scaling regime in theories with a gap in their excitation spectrum and trivial, ie. the only relevant operator around the Gaussian fixed point is the quadratic $\phi^2$. We shall see that $\phi_a$ may jump during the evolution and there might be several IR scaling regimes along some renormalized trajectories.

\subsection{Strong renormalization close to the phase boundary}
\begin{figure}
\includegraphics[width=5cm]{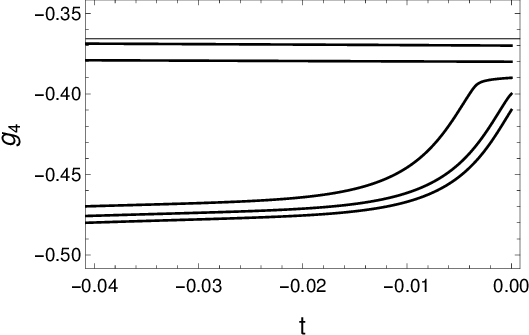}\hskip1cm
\includegraphics[width=5cm]{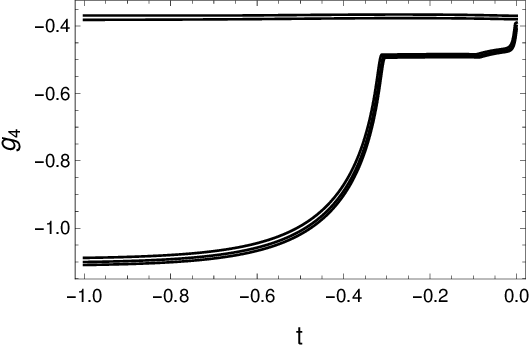}

\hskip1.cm(a)\hskip6cm(b)

\includegraphics[width=5cm]{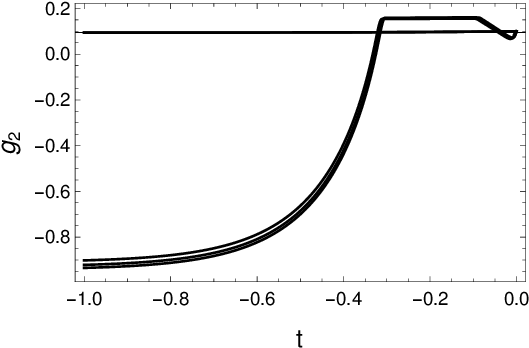}\hskip1cm
\includegraphics[width=5cm]{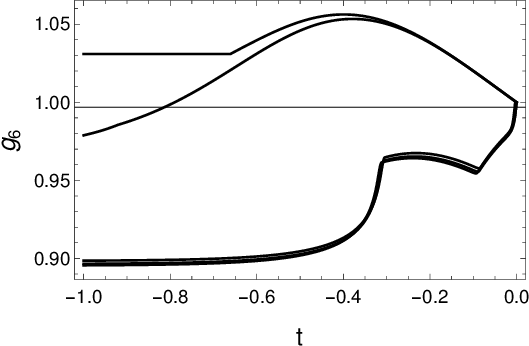}

\hskip1.cm(c)\hskip6cm(d)
\caption{The evolution of the parameters along the first order phase boundary for $g_{B2}=0.1$,  $g_{B4}=-0.38,-0.39,-0.4,-0.41,-0.42$, and $g_{B6}=1$. (a): $g_4$ close to the initial cutoff, (b): $g_4$, (c): $g_2$  and (d): $g_6$ following a longer evolution.}\label{fivetrajf}
\end{figure}

The absolute minimum of the potential remains either vanishing $\phi_a=0$ or non-vanishing $\phi_a=\phi_m$ all along the renormalized trajectories deeply within the symmetric or the symmetry broken phase, respectively. But contrary to continuous phase transitions $\phi_a$ may jump between 0 and $\phi_m$ during the evolution close to the first order phase transition. The corresponding jump of $G_2(\phi_a)$ may throw the system from one scaling regime to another rendering the global features of the renormalized trajectory, in particular the relation between the UV and the IR parameters, becomes highly involved.

This mechanism is demonstrated in Fig. \ref{fivetrajf} where the evolution of the parameters are shown close to the separatrix of the first order transition for different $g_{B4}$ with fixed $g_{B2}$ and $g_{B6}$. The tree-level transition line is at $g_{B4}=g_{4tr}=-0.4$ and the trajectories with initial value $g_{B4}=-0.38,-0.39,-0.4,-0.41,-0.42$ are displayed. The evolution of $g_4$ is weak and smooth for $g_{B4}>-0.4$ in the symmetric phase where $\phi_a$ remains vanishing according to Fig. \ref{fivetrajf}(a). But a violent renormalization sets in slightly below $g_{B4}=-0.4$ because the beta functions change in a discontinuous manner at $g_{B4}=-0.4$. Thus the slightest renormalization of $g_2$ and $g_6$ just below $g_{B4}=-0.4$ may induce a jump of $\phi_a$ along the trajectory. Actually $\beta_2(\phi_m)$ is slightly positive and $\beta_6(\phi_m)$ is very close to zero at $k\sim k_i=1$ hence the tree level transition point $g_{4tr}$ is increased by radiative corrections and the trajectory with $g_{B4}=-0.4$ is actually driven by $\beta_2(\phi_m)$ startig with the first infinitesimal $k\to k-\dk$ step. As a result the separatrix is pushed towards slightly larger $g_4$ values compared to the tree-level solution.

The beta functions at $\phi_a=\phi_m$ tend to be larger in absolute magnitude than the those at $\phi_a=0$ and when $\phi_a$ jumps and the evolution slows down (speeds up) at the jump $\phi_a=\phi_m\to0$ ($\phi_a=0\to\phi_m$). The renormalization becomes more violent immediately after a jump $\phi_a=0\to\phi_m$ owing to the larger beta functions. However the jump throws the system into an IR scaling regime where the renormalization slows down. Such a phenomenon is well recognizable on the trajectoires of Figs. \ref{fivetrajf} belonging to the symmetry broken phase where $\phi_a$ jumps twice along the trajectories and $\phi_a=0$ for $-0.33<t<-0.08$.

\begin{figure}
\includegraphics[width=5cm,angle=-90]{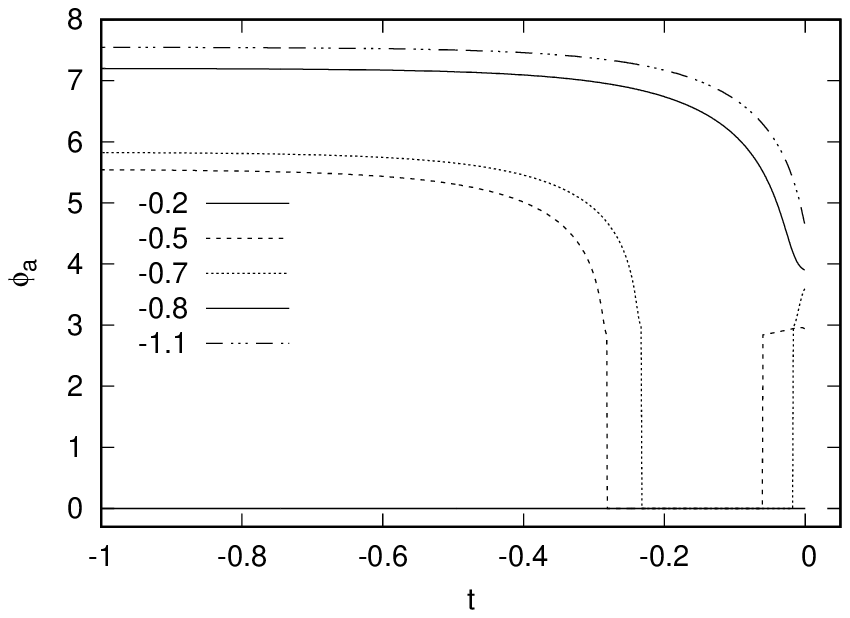}\hskip1cm
\includegraphics[width=5cm,angle=-90]{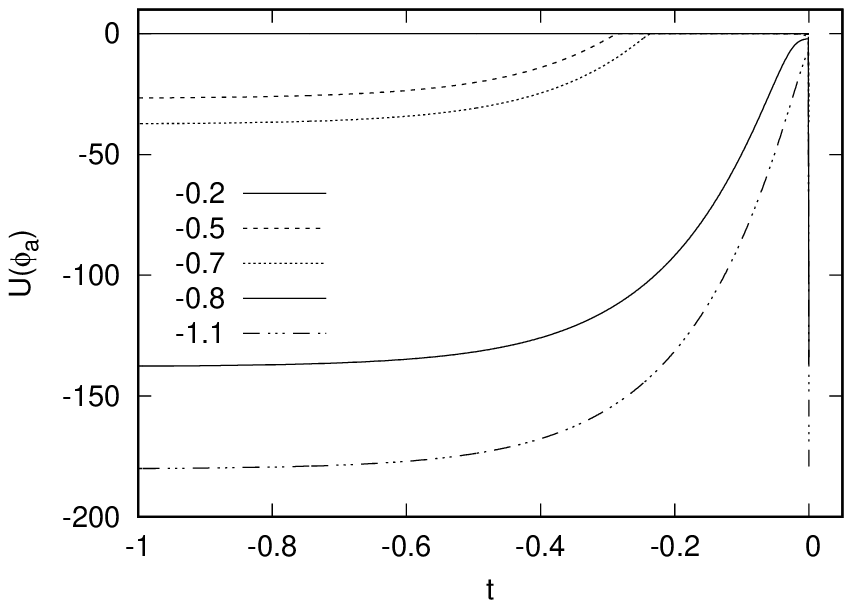}

\hskip.2cm(a)\hskip8cm(b)

\includegraphics[width=5cm,angle=-90]{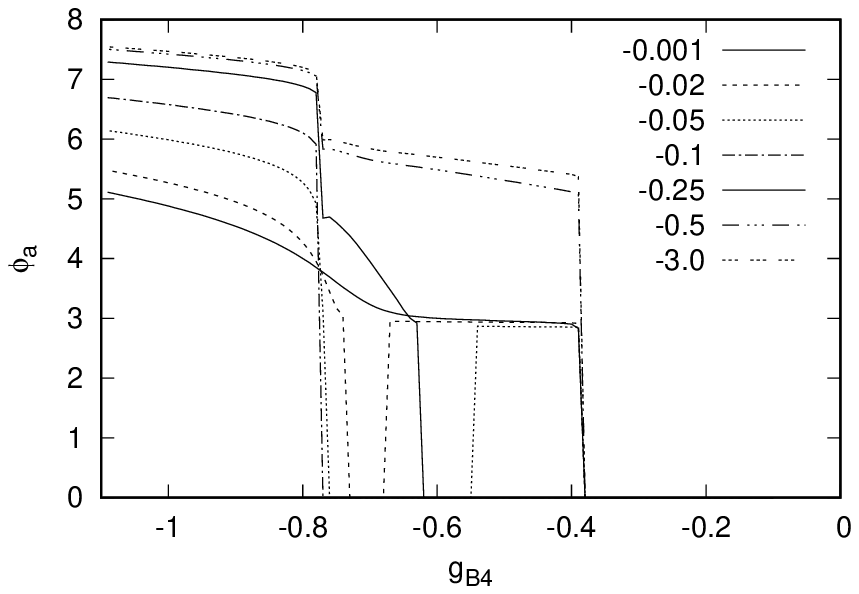}

\centerline{(c)}
\caption{The location of the absolute minimum and the value of the potential for $g_{B2}=0.1$, $g_{B6}=1$. (a): $\phi_a$, (b): $U(\phi_a)$ as functions of the gliding scale parameter $t$. The values of $g_{B4}$ are in the inset. (c): $\phi_a$ as the function of $g_{B4}$, read off at different scale parameter $t$, shown in the figure.}\label{typtrajf}
\end{figure}

The evolution of $\phi_a$ and the value of the potential $U(\phi_a)$ are shown in Figs. \ref{typtrajf}(a) and (b) for few typical trajectories for the same initial value of $g_{B2}$ and $g_{B6}$ and different choice of $g_{B4}$. Deeply within the symmetrical ($g_{B2}=-0.2$) and the symmetry broken phase ($g_{B2}=-0.8,-1.1$) $\phi_a$ starts and stays at $\phi_a=0$ and $\phi_a\ne0$, respectively, as expected. These trajectories constitute the picture of the first order phase transition, found in earlier works \cite{tetradisw,teradis,jakubszyk}, where the competition of the minima was not followed. However the minimum $\phi_a=\phi_m$ falls back to $\phi_a=0$ in our treatment within a finite scale interval $t_1<t<t_2$ during the evolution within the symmetry broken phase closer to the transition, at $g_{B4}=-0.5$ and $-0.7$.

It is important to keep in mind that the trajectories remain an analytic function of the initial conditions and of the scale $k$ for $0<k<k_i<\infty$ owing to the continuity of the weighted average \eq{expmix} with $c_s<\infty$. The discontinuity of $\phi_a$ in Fig. \ref{typtrajf}(a) is a mathematical artefact without relevance for physics.

\subsection{Locking into a degenerate potential}\label{lockings}
What is rather remarkable on Fig. \ref{typtrajf}(b) that the potential remains degenerate in a finite scale interval $0<t_2-t_1<\infty$. Closer inspection of these trajectories reveals an oscillatory behaviour: When $\phi_a=0$ then the simple the beta functions evolve the potential in such a manner that $U(\phi_m)>0$ decreases and until it becomes negative. From that on the complicated beta functions are used at $\phi_m$ and they push $U(\phi_m)$ back to positive values and the cycle starts again. Such an oscillation can be understood by the linearization of the beta functions around the degenerate values $g_{nd}$,
\be
\Delta\dot g_n=\Theta_{c_s}(\Delta g_n)\Delta\beta_n^{(+)}+\Theta_{c_s}(-\Delta g_n)\Delta\beta_n^{(-)}
\ee
where $\Delta g_n=g_n-g_{nd}$ is the deviation from the degeneracy $g_4=g_{4tr}$, $\Theta_{c_s}(x)$ is the Heaviside function smeared in an $\ord{c_s^{-1}}$ interval around zero and $\Delta\beta_n^{(\pm)}$ are  continuous functions of the parameters. Such an $\ord\dk$ oscillation of the renormalized trajectory appears to be stable against the change of the parameter $c_s$ and is supported only in a limited scale interval $t_1<t<t_2$ where $\pm\Delta\beta_n^{(\pm)}>0$. The oscillation starts when the inequality sets in and stops when the ongoing renormalization leads to its violation.

Such a locking into a degenerate potential is important because the size $t_2-t_1$ of the scale window was found to be a discontinuous function of the initial condition, $g_{Bn}$. Therefore the locking in mechanism generates a discontinuity between the UV and the IR parameters, namely a first order phase transition. This is shown in Fig. \ref{typtrajf}(c) where $\phi_a$ is plotted against $g_{B4}$ at different scale $k$. The evolution is weak in the IR regime and $\phi_a$ at $t=-3$ can already qualitatively be treated as the order parameter which is the location of the absolute minimum at $k=0$. The curves display two jumps in the IR scaling regime for $t<-0.5$, one around $g_{B4}=-0.38$ and another around $g_{B4}=-0.76$. The jump around $g_{B4}=-0.38$ is due to the separatrix seen in Figs. \ref{fivetrajf}. The emergence of the second first order transition is due to the abrupt disappearance of the locking mechanism.

The locking of the renormalized trajectory into a degenerate potential is reminiscent of the phase mixing at the transition point. In fact, one expects that a droplet changes the dynamics when the cutoff is between  size of the domain wall and the full domain. However we tend to disregard the identification of the phase coexisting region and the interval between the two fist order transitions because the former joins the two phases without first order transitions at the edges.

The strong evolution within the symmetry broken phase but close to the phase boundary generates a complicated scale dependence for $\phi_a$ for $t\sim0$ around $g_{B4}=g_{tr4}=-0.4$. In fact, $\phi_a=\phi_m$ during a very short evolution the but it jumps to zero on some trajectories before $t=-0.02$. The length of the scale window with $\phi_0=0$ is increasing during the evolution but disappears before we reach $t=-0.25$. The two jumps stabilizes its position in the final IR scaling regime $t<-0.5$.

It is worthwhile noting that the weak dependence of the location of the jumps of $\phi_a$ on the scale $k$, seen in Fig. \ref{fivetrajf}(c), is not an indication of weak dressing, a slow evolution along the renormalized trajectories as in the case of continuous phase transitions. What happens here is that the discrete jump of $\phi_a$ taking place at a certain value of $k$ leaves a non-recoverable impact on the trajectories and remains to be felt down to $k=0$.

\subsection{Quasiparticle mass}
The running parameter $G_2$ can be interpreted with a slight abuse of the language as the mass square of the quasiparticles at the actual scale $k$. How does the running quasiparticle mass evolve compared to the  initial cutoff $k_i$? We see the tree-level value of $\phi_a$ at the initial conditions and the mass changes by a factor two as we pass the case of degenerate initial potential. Since the location of this discontinuity is shifted very weakly there seems to be no problem to place the phase transition separating the symmetric and the symmetry broken phases at any mass at the initial condition in the symmetry broken phase.

\begin{figure}
\includegraphics[width=5cm,angle=-90]{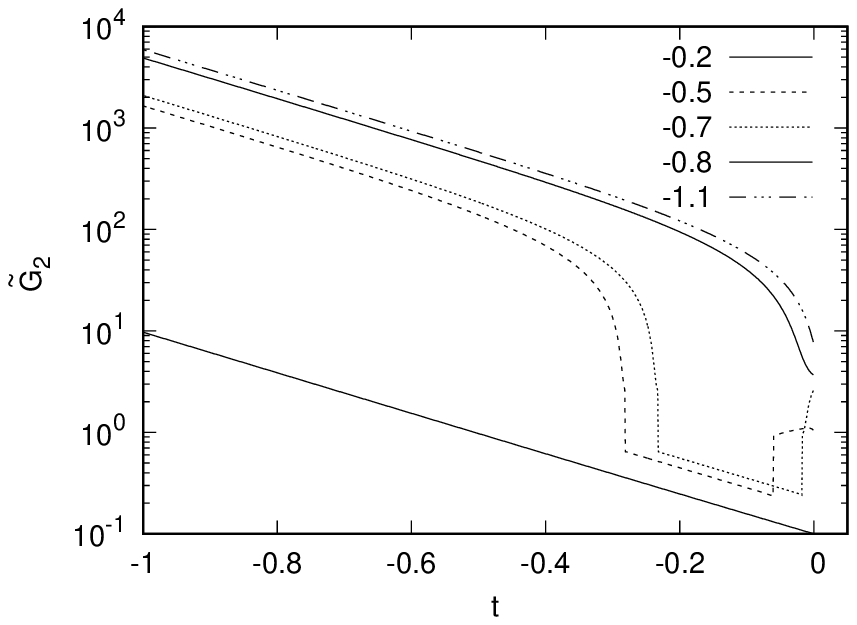}\hskip1cm
\includegraphics[width=5cm,angle=-90]{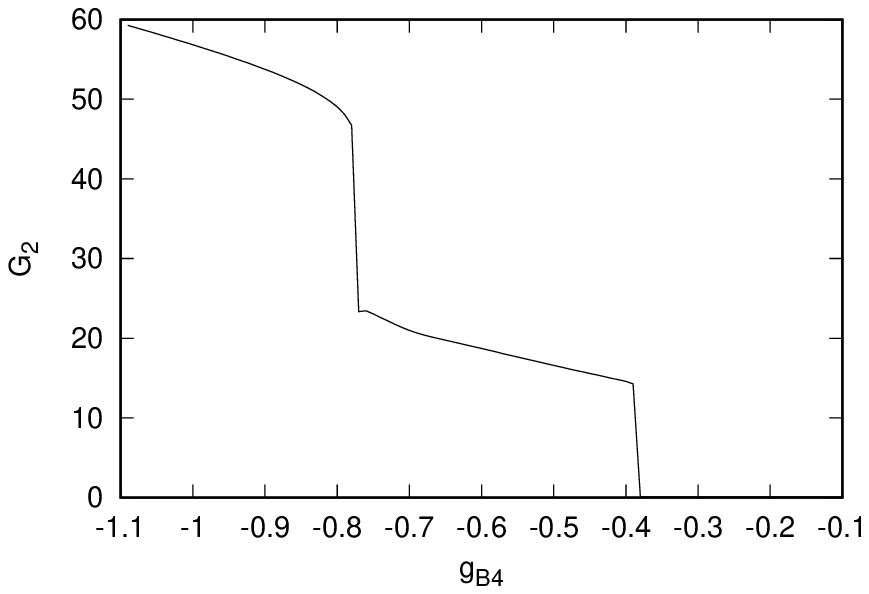}

\hskip.2cm(a)\hskip8cm(b)
\caption{(a): The evolution of $\tilde G_2$ and (b): $G_2$ deep in the IR at $t=-3$ as the functions of $g_{B4}$ for the same value of $g_{B2}$ and $g_{B6}$ as in Fig. \ref{typtrajf}.}\label{mass}
\end{figure}

However this picture is changed considerably according to Figs. \ref{mass}. The mass square $G_2$ remains remarkably close to it tree-level value, $g_{B2}=0.1$ with weak evolution along the trajectory $g_{B4}=-0.2$ in the symmetric phase where the UV-IR crossover is around $t=-0.4$ according to Fig. \ref{mass}(a). But the trajectories in the symmetry broken phase suffer strong renormalization. In fact, the trajectories undergo a strong renormalization at the very beginning of the evolution close to $g_{4B}=g_{4tr}=-0.4$ decreasing quickly the running mass. The mass stays approximately constant within the locking in scale window but is pushed up again by the returning of the $\phi_a=\phi_m$ scaling laws, governing the evolution from the rest of the trajectory. The IR values of $G_2$ is reproduced in Fig. \ref{mass}(b) shows clearly that the symmetry broken theory is fully in the IR scaling regime without UV scaling laws apart of the finite locking in scale interval which exists close to the phase boundary.

Hence the free choice of the initial mass is strongly overwritten by the strong renormalization at $\phi=\phi_m$. Actually no light mass symmetry broken phase, $G_2/k^2_i\ll1$, was found numerically.

\subsection{Second order transition}
We have so far sought first order transitions driven by sufficiently negative $g_{B4}$ with $g_{B2},g_{B6}>0$. But there might be second order transitions for less negative $g_{B4}$, as well. The second order spontaneous symmetry breaking of the Ising model universality class in the traditional $\phi^4$ model is the result of large fluctuations around the trivial vacuum, $\phi=0$, supported by a double well bare potential. The potential of the $\phi^6$ model has local minima at $\phi_m$ in the symmetric phase for $-\sqrt{8g_2g_6/5}<g_4<-\sqrt{6g_2g_6/5}$ around which finite life-time quasiparticles can be formed. For low enough tunnelling probability to the trivial minimum the life-time might be long enough to induce a second order phase transition.

\begin{figure}
\includegraphics[width=5cm,angle=-90]{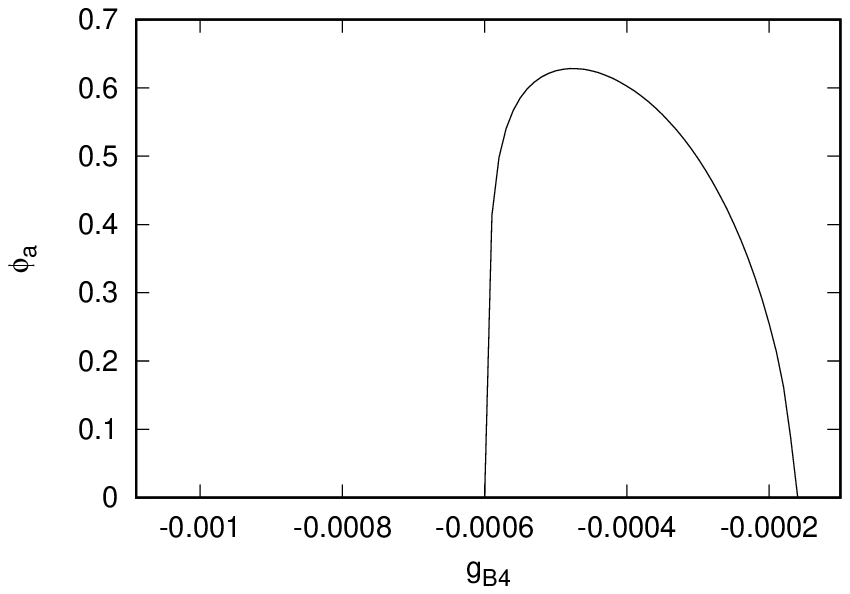}
\caption{The location of the absolute minimum, $\phi_a$, deep in the IR at $t=-3$ as the functions of $g_{B4}$. The other bare parameters are $g_{B2}=10^{-6}$ and $g_{B6}=10^{-2}$.}\label{secondf}
\end{figure}

A second order phase transition is indeed found in the $\phi^6$ model as indicated by Fig. \ref{secondf} where  $\phi_a$ is plotted as the function of $g_{B4}$ for $g_{2B},g_{B6}>0$. We are in the symmetric phase at $g_{B4}=0$ and the moving of $g_{B4}$ in the negative direction makes $U(\phi_m)$ smaller, the excitations around it more stable which facilitates a second order spontaneous symmetry breaking. The location of the non-trivial minimum, $\phi_m$, is increased during the decrease of $g_{B4}$ and the increasing tunnelling factor brings ultimately the system back to the vicinity of the energetically favorable $\phi=0$ by restoring the symmetry. The result is a narrows strip of symmetry broken phase within the symmetric region.

But such an explanation is very naive and can not be taken more than an educated guess. In fact, $g_{4tr}=10^{-4}\sqrt{8/5}$ for the parameters of the figure hence fluctuations must be very strong to push the first order phase transition away from $g_{4tr}$ and invalidate the tree-level estimate. The first order transition was found at more negative $g_{B4}$ but a much more detailed analysis is needed to find out whether the first and second order transitions for $g_{B2},g_{B6}>0$, $g_{B4}<0$ and $g_{B4},g_{B6}>0$, $g_{B2}<0$, respectively match and together separate the symmetric and the symmetry broken phases or there is a region connecting the two phases without phase transition.

\subsection{Alternative signature of spontaneous symmetry breaking}
The usual test of the ferromagnetic phase is to apply an external magnetic field and to follow the magnetization as the external field is decreased. The non-vanishing limit of the magnetization when the external field is removed is the signature of the ferromagnetic phase. Naturally the first order phase transition seen by scanning in the external magnetic field is not identical, only a related, to the second order phase transition in the temperature.

\begin{figure}
\includegraphics[width=5cm,angle=-90]{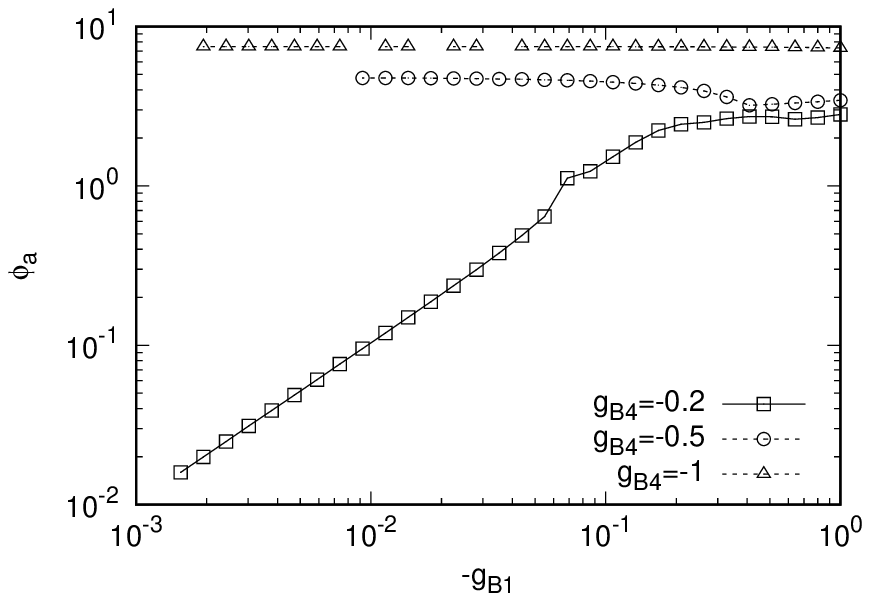}\hskip1cm
\includegraphics[width=5cm,angle=-90]{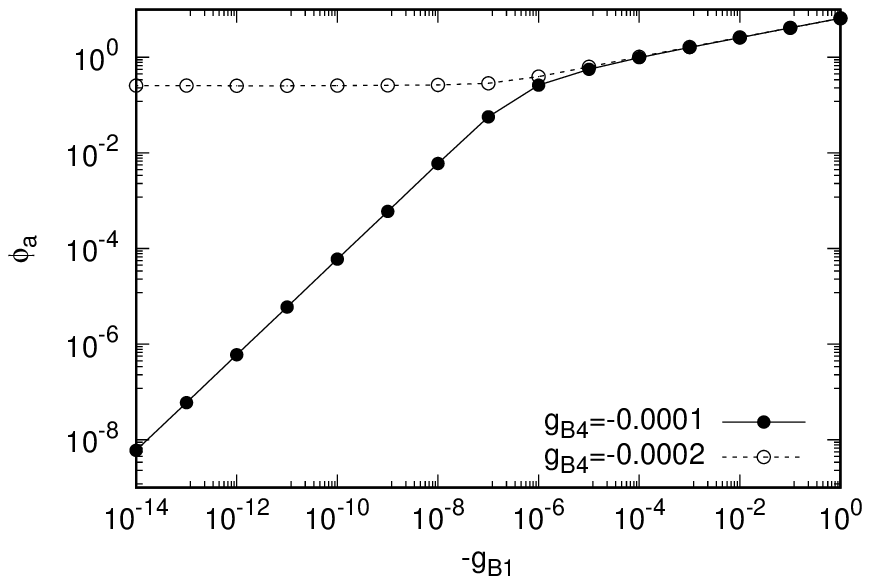}

\centerline{(a)\hskip8cm(b)}
\caption{The restoration of the symmetry as $g_{B1}\to 0$. (a): The first order phase transition of Fig. \ref{typtrajf}. (b): The second order phase transition of Fig. \ref{secondf}.}\label{g1jumpf}
\end{figure}

Thereby we can corroborated the phase structure by introducing a linear term in the potential,  $U(\phi)\to U(\phi)+g_1\phi$, and by scanning the dependence of $\phi_a$ on $g_{B1}$. One expects a continuous $g_{B1}$-dependence within the symmetric phase and a discontinuity at $g_{B1}=0$ in the phase with spontaneous symmetry breaking. The absolute minimum $\phi_m$ as the function of $g_1$, shown on the two sides of the first and the second order transition line in Figs. \ref{g1jumpf} affirms this behaviour. Note that the latent heat is vanishing in these first order phase transitions.

\section{Lattice regularization}\label{lattices}
The impressive richness of the phase structure, produced by radiative corrections, raises the question of reliability and calls for an independent check. In fact, the radiative corrections may substantially be modified by our approximation, the restriction of the blocked action space, and the choice of the subtraction procedure. A natural way to check the phase structure is to calculate the order parameter by a Monte-Carlo simulation of a lattice regulated version of the theory.

The bare action is given by the space-time sum
\be
S[\phi]=\sum_n\left[\hf\sum_{j=1}^3(\phi_{n+\hat j}-\phi_n)^2+g_{L1}\phi_n+\frac{g_{L2}}2\phi_n^2+\frac{g_{L4}}{4!}\phi_n^4+\frac{g_{L6}}{6!}\phi_n^6\right]
\ee
for lattice spacing $a=1$ where vector $n=(n_1,n_2,n_3)$ with $n_j=1,\ldots,N_L$ labels the lattice sites of an $N_L^3$ lattice equipped with periodic boundary conditions, $\hat j$ denotes the unit vector in the direction $j$ and the partition function is given by the integral
\be\label{partf}
Z=\prod_n\int d\phi_ne^{-S}.
\ee
A sweep of the Monte-Carlo update consists of bringing the local field into contact with a heat bath over the whole lattice in a sequential manner. The heat bath was a particular realization of the Metropolis algorithm where a shift, $\phi_n\to\phi_n+\Delta\phi$ distributed homogeneously in the interval $-\chi<\Delta\phi<\chi$, was offered to the field variable. This change was accepted with the probability $\min(1,r)$ where $r$ denotes the ratio of the integrand of the partition function with the shifted and the original field value. This process was repeated $n_M=6$ times  before moving to the next lattice site and the parameter $\chi$ was chosen in such a manner that the acceptance ratio averaged over a sweep stayed within the interval $[0.45,0.55]$.

The path integral \eq{partf} has two regulators, the UV and the IR cutoffs, the lattice spacing $a$ and the lattice size $N_L$, respectively and they leave important differences compared to the partition function \eq{partfnct} in the continuous space-time. We intend to compare observables obtained by different regularizations thus one should determine first the bare parameters of the theory by matching some observables obtained in both calculations. This is beyond the scope of this work where we content ourself to compare the qualitative features of the phase structure.

To minimize the lattice artefacts the correlation length in lattice spacing units $\xi/a$ should be large.
The correlation lengths, the Compton wavelength of the lightest particle created by the local field, is found by projecting the field variable onto to the vanishing spatial momentum sector, $\varphi_{n_1}=\sum_{n_2,n_3}\phi_n$ and calculating the connected correlation function
\be\label{conncf}
G^{(0)}_{n_1-n'_1}=\la\varphi_{n_1}\varphi_{n'_1}\ra-\la\varphi_{n_1}\ra\la\varphi_{n'_1}\ra.
\ee
The asymptotic decay $G^{(0)}_{n_1}\approx ce^{-man_1}$ for large $n_1$ defines the lightest mass, the inverse Compton wavelength in units of the UV cutoff, $ma=a/\xi$.

The numerical calculation are restricted to a finite volume where there are no phase transitions. Thus the lattice should be large enough to avoid the symmetry restoration by the flip-flops of the slow mode, the sudden changes of sign of the order parameter
\be
\Phi=\frac1{N_L^3}\sum_n\phi_n
\ee
during the simulaton. The lattice size $N_L=100$ was used in the numerical work and no flip-flop was seen.

\subsection{Hysteresis cycles}
\begin{figure}
\includegraphics[width=5cm,angle=-90]{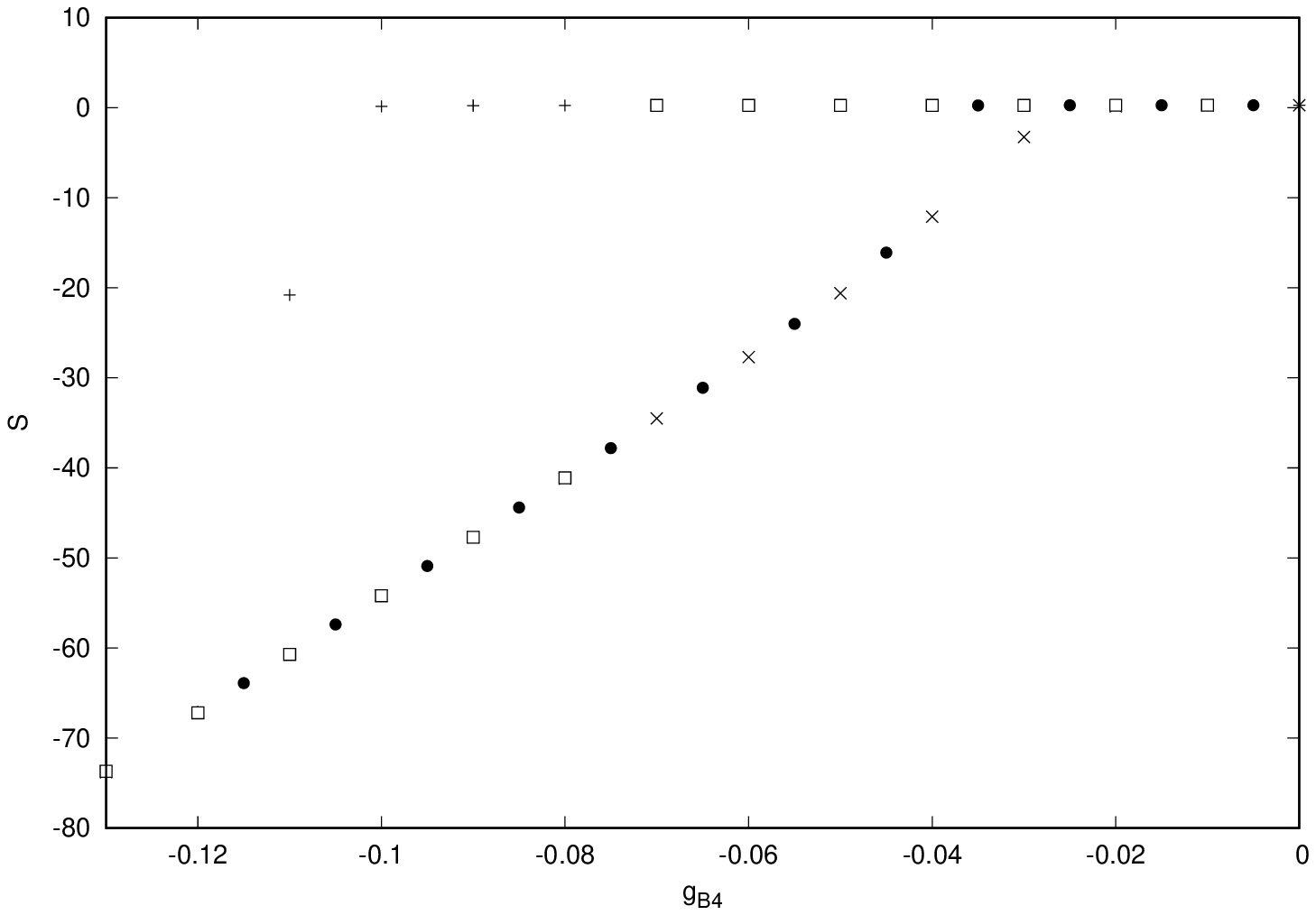}\hskip1cm
\includegraphics[width=5cm,angle=-90]{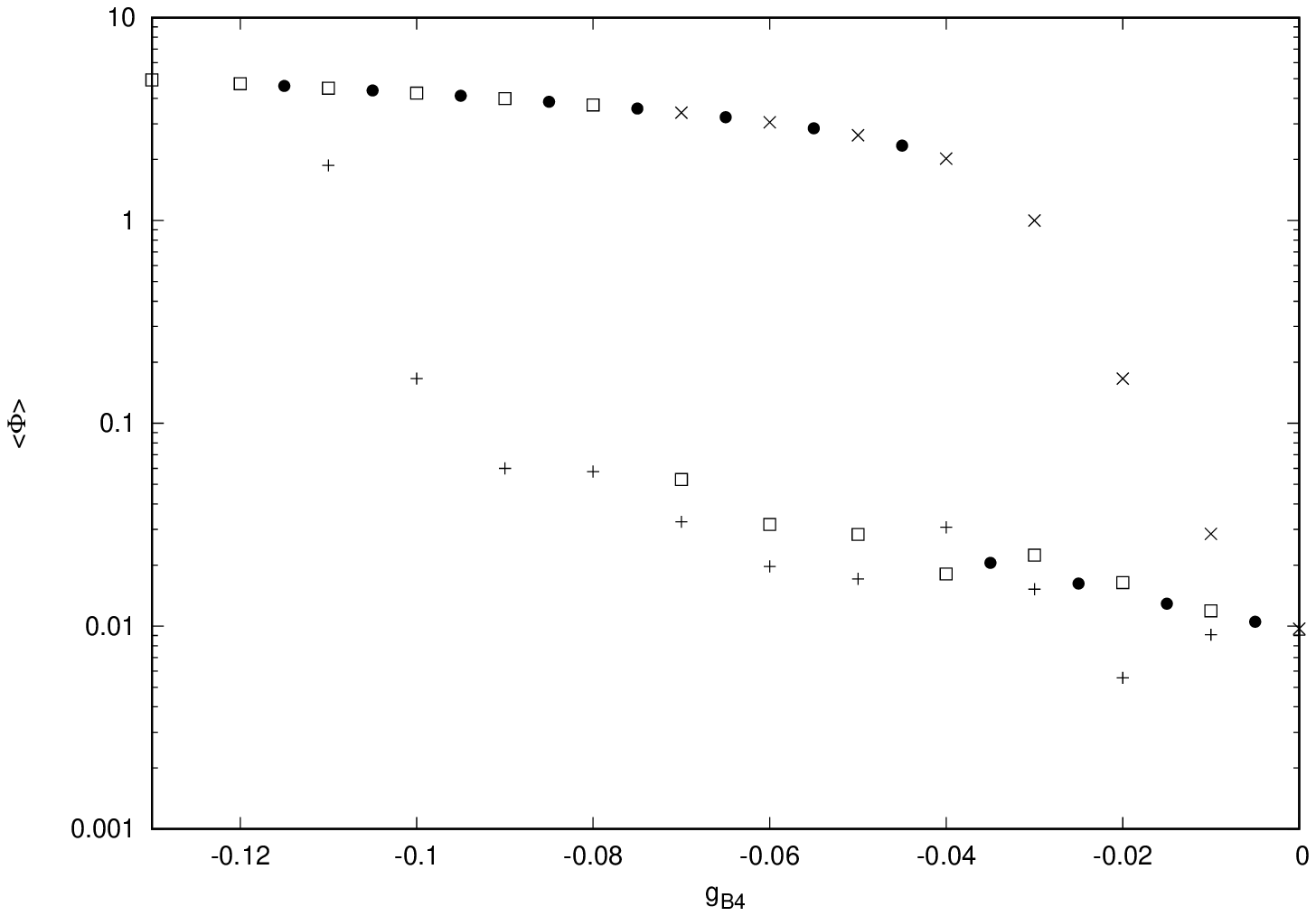}

\includegraphics[width=5cm,angle=-90]{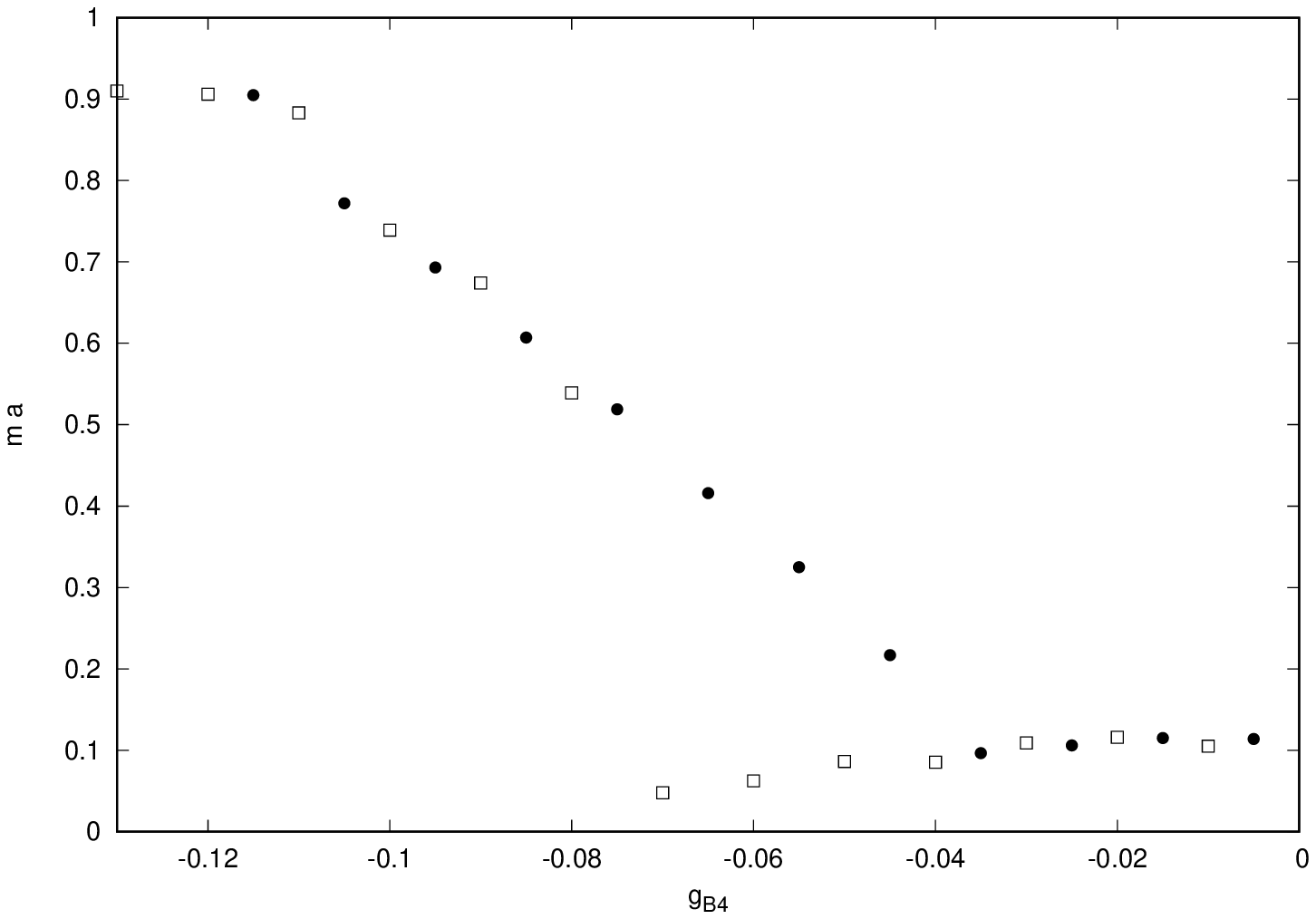}
\caption{The hysteresis cycle for the action $S$, the order parameter $\Phi$, and the inverse mass $1/ma$ in the coupling constant $g_{L4}$ with $g_{L2}=0.01$ and $g_{L6}=0.1$. The symbols $+$ ($\Box$) and $\times$ ($\bullet$) indicate the down and the up moving part of the cycle with 1000 (at most 250000) sweeps at each point. The inverse mass is shown only for the longer iteration series.}\label{histf}
\end{figure}

In order to check the convergence of the Monte-Carlo iterations two hysteresis cycles were made in $g_{L4}$ with the results shown in Figs. \ref{histf}. The simulation started with an ordered configuration $\phi_n=0$ and $g_{L4}$ was moved through the values $g_{L4}=-0.01~n$, $n=0,\ldots,14$ by executing $1000$ sweeps with fixed $g_{L2}=0.01$, $g_{L6}=0.1$ in such a manner that the last configuration at a $g_{L4}$ value was used as the initial configuration for the next $g_{L4}$ point. At the end the final configuration was taken and the values $g_{L4}=-0.115+0.01~n$ were visited in an opposite move in a similar sequential manner. Another hysteresis cycle was performed with at most $250000$ sweeps at the same $g_{L4}$ values but in this case the initial configuration was $\phi_n=0$ and $5$ at $g_{L4}=-0.01~n$ and  $g_{L4}=-0.115+0.01~n$, respectively. The iterations stopped when the order parameter converged in the last $40000$ sweeps in the energetically preferred phase. This rearrangement allows to see the speed of convergence of the simulation.

The action $S$ and the order parameter $\Phi$, depicted in Figs. \ref{histf}(a)-(b), represent an UV and an IR observable with a much faster convergence for the former than for the latter. The approximately linear dependence of the action on $g_{L4}$ can be understood by assuming that the fluctuations are weak at the scale of the lattice spacing since the expectation value of the order parameter is approximately $g_{L4}$ independent within this range according to Fig. \ref{histf}(b).

The first order phase transition behind the hysteresis cycle Figs. \ref{histf} separates the disordered ($\la\Phi\ra=0$) and the ordered ($\la\Phi\ra\ne0$) phases. Another first order transition in this theory is between the two ordered vacua, $\la\Phi\ra=\pm\Phi_0$, its signature is the hysteresis cycle plotted in Fig. \eq{phig1hf}. Here $g_{L1}$ was moved through the interval $-0.3<g_{L1}<0.3$ from left to right in the symmetry broken phase in a sequential manner as in the case of the previous hysteresis cycle with $1000$ iterations. In the other half of the cycle, not shown in the figure, one finds the same curve with the transformation $g_{L1}\to-g_{L1}$.

\begin{figure}
\includegraphics[width=5cm,angle=-90]{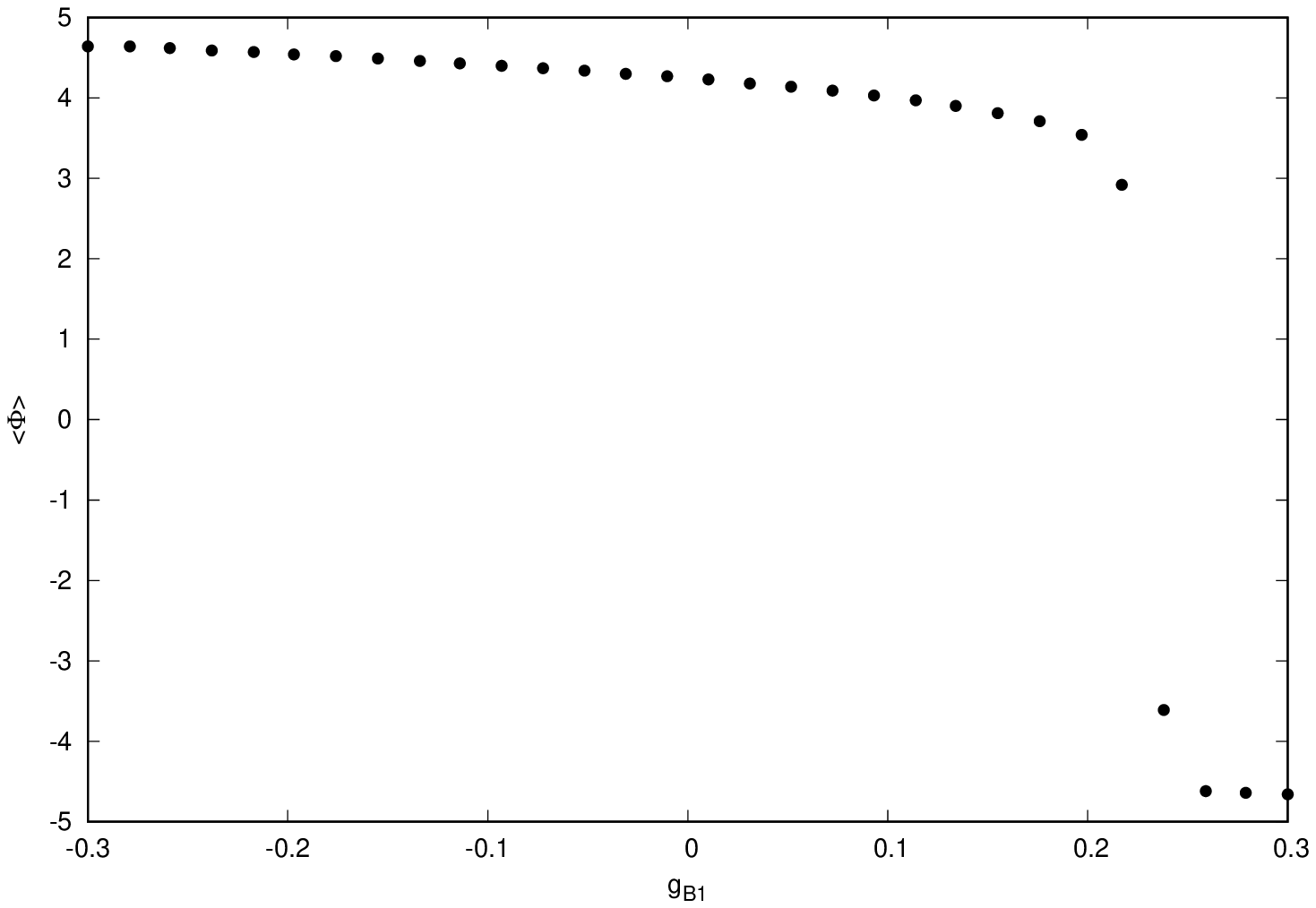}
\caption{Half of the hysteresis cycle of the order parameter in $g_{L1}$ with the bare parameters are $g_{L2}=0.01$, $g_{L4}=0$ and $g_{L6}=0.1$.}\label{phig1hf}
\end{figure}

\subsection{Limitation in the UV}
\begin{figure}
\includegraphics[width=5cm,angle=-90]{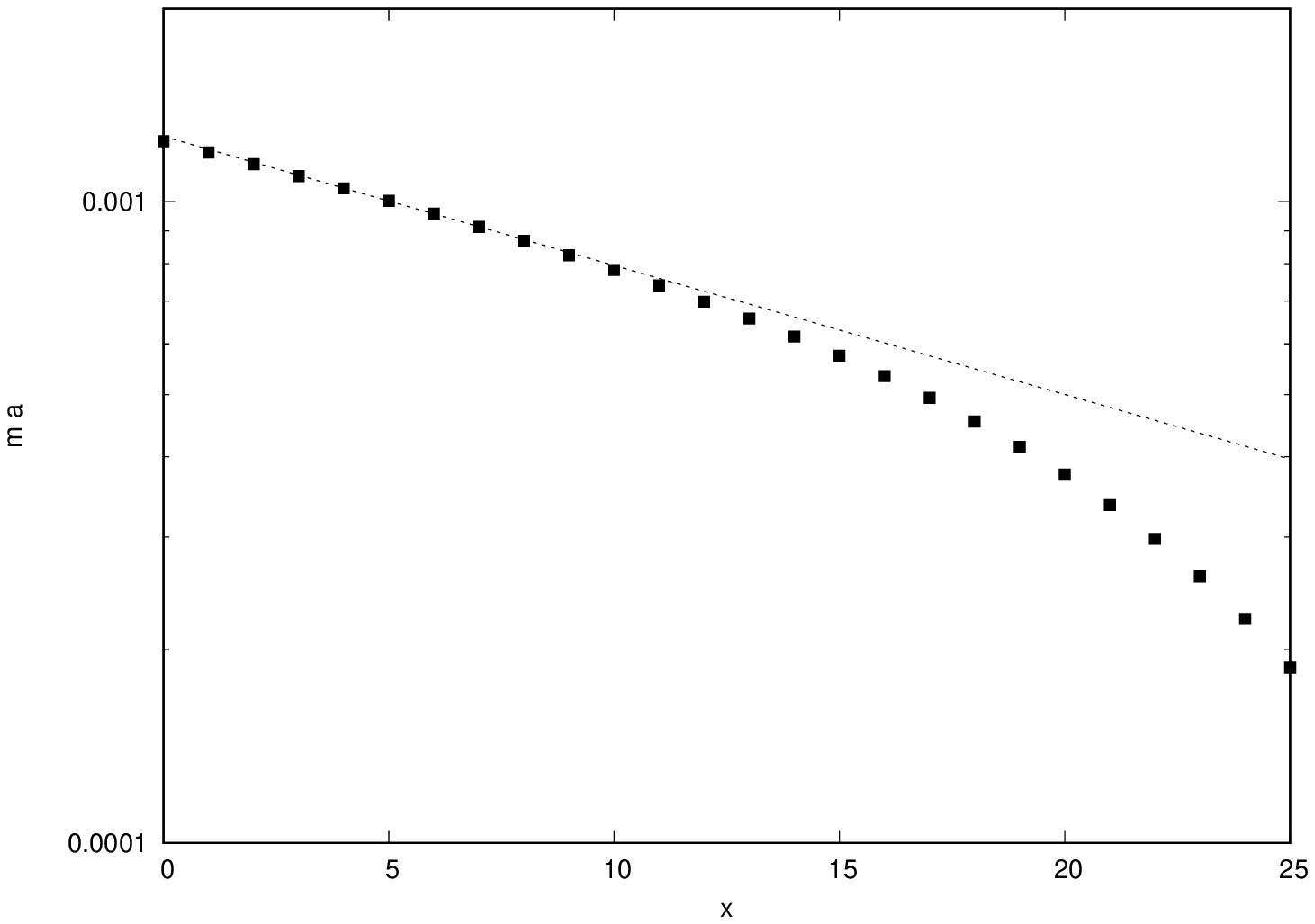}
\caption{The connected $\varphi_{n1}$ correlation function denoted by squares and its fit indicated by dotted line for $g_{L2}=0.01$, $g_{L4}=-0.07$, and $g_{L6}=0.1$. The best estimate of the error bar is the difference between the average at $n_1=25$ and the exponential fit. The figure is based on the data set corresponding to the last 200 points of the series shown in Fig. \ref{phievolf}.}\label{gapg4f}
\end{figure}

The rich phase structure predicted by the renormalization group method is due to the strong renormalization. To recover the fine balance between the diferent contributions to the beta functions one has to suppress the lattice artefacts of the free lattice propagator, achieved by performing the continuum limit $\xi=1/ma\to\infty$ diverges. The calculation of the mass gap is shown for $g_{L4}=-0.07$ in Fig. \ref{gapg4f}, the result of a fit of the logarithm of the connected correlation function \eq{conncf} by a linear function of the distance. The mass gap, obtained by such fitting, is shown for the full hysteresis cycle on Fig. \ref{histf}(c). The statistical errors are smaller than the symbol size and the systematic errors can be estimated by the difference between the two hysteresis series.

The lesson is that we are far from the continuum limit when the order parameter is non-vanishing. One could reduce the mass by going closer to the Gaussian fixed point but a large amplitude, slow wandering of the order parameter creates a serious numerical problems. Such a critical slowing down close to a second order phase transition can in principle be reduced by Fourier acceleration of the slow modes \cite{batrouni}, multigrid \cite{goodman} and cluster update \cite{swendsen}.

\subsection{Limitation in the IR}\label{irlims}
A fully converged simulation produces no hysteresis cycle which arises from the simulatoin getting stuck in the false vacuum. Thus Fig. \ref{histf} and \ref{phig1hf} report a serious slow down of the convergence  around first order phase transitions, a well known limitation of the Monte-Carlo method.

\begin{figure}
\includegraphics[width=5cm,angle=-90]{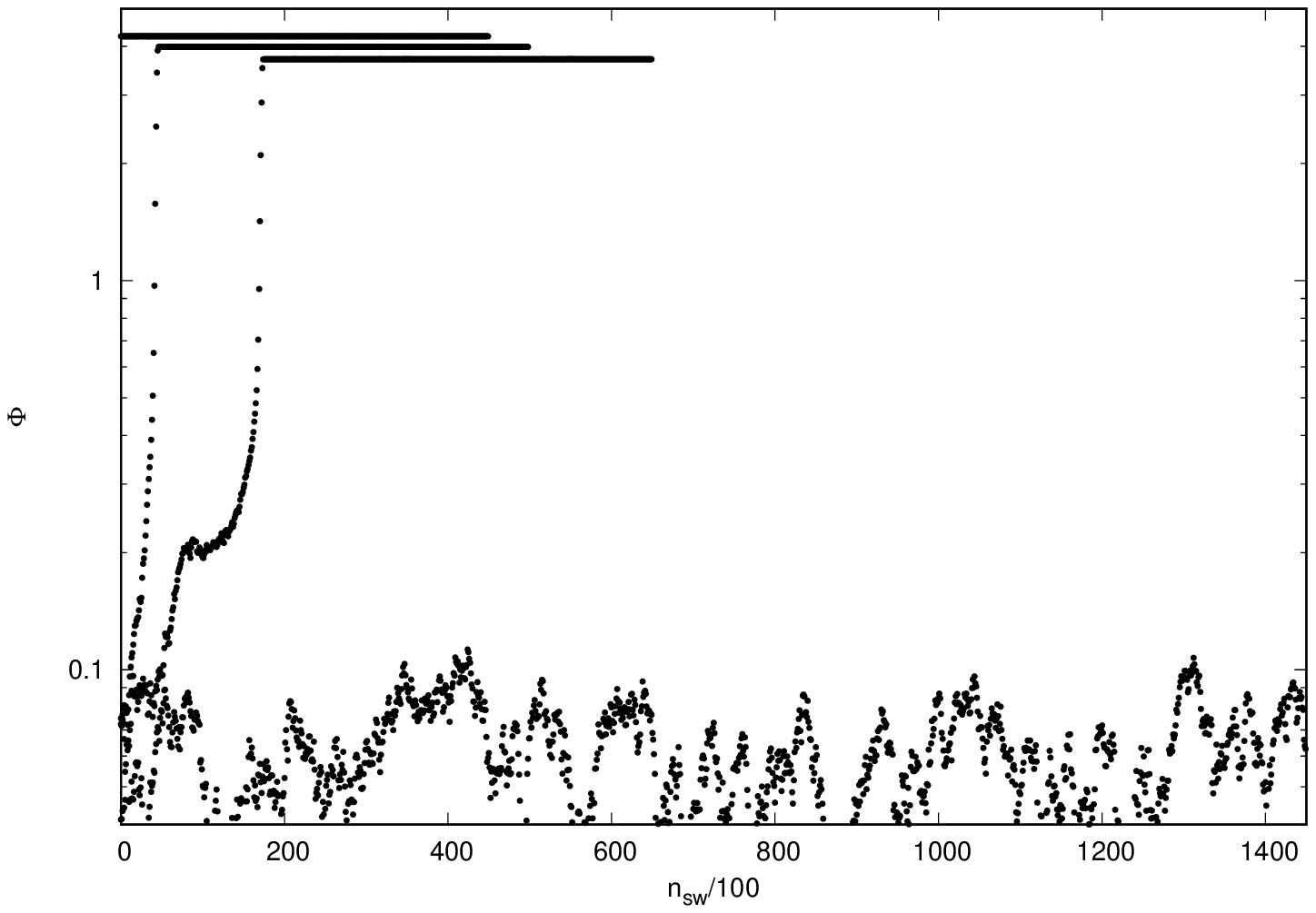}
\caption{The Monte-Carlo sequences of the order parameter. A point corresponds to the absolute magnitude of the order parameter averaged over 100 sweeps. The bare parameters are $g_{L2}=0.01$, $g_{L6}=0.1$ and $g_{L4}=-0.07,-0.08,-0.09$ and $-0.1$. The order parameter increases as $g_{L4}$ moves towards the negative direction.}\label{phievolf}
\end{figure}

The point $\phi=0$ remains a local minimum of the bare potential for any $g_{L4}$. However the finite length Monte-Carlo iteration series lead to the stable vacuum with non-vanishing order parameter only for sufficiently negative $g_{L4}$, beyond the true transition point. A few typical Monte-Carlo series are shown in Fig. \ref{phievolf}, they represent three points on Fig. \ref{histf}(b). The simulation time needed to find the energetically stable true vacuum is supposed to increase exponentially with the volume within the interval $-0.09<g_{L4}<-0.07$. Once the stable vacuum is reached the system stays there. There is a similar problem to find the transition by starting from the ordered phase with non-vanishing order parameter. It is interesting that this part of the cycle with more iterations closes at the tree-level transition point at $g_{L4}=g_{4tr}=-0.04$. This might be an accident since the renormalization between the lattice spacing and the size of the lattice is very strong, see section \ref{sidteresiss} below. The lesson of Fig. \ref{phig1hf} is similar since the simulation with finite iteration series find the transition only when the true vacuum is much below the false one.

The hysteresis cycle of Fig. \ref{typtrajf} covers the region where one expects the second first order transition predicted by the renormalization group method. as far as the second order phase transitions of Fig. \ref{secondf} are concerned, one needs small bare parameters. However the order parameter follows a slow but large amplitude oscillation during the simulation in the symmetric phase because the restoring force at $\Phi=0$, $U'(0)$, is very weak and one encounters a slow down similar to the one arising in the continuum limit.

\subsection{Droplets}\label{sidteresiss}
The droplets are supposed to provide the driving force of a first order transition \cite{langer,fischer,voloshi,binder} hence their presence is crucial in the Monte-Carlo simulation. Their identification during the iteration serves two goals in the same time. First, it may help to find the phase coexisting region, a problem raised in section \ref{lockings}, and it might help to design improved update schemes with faster convergence.

It is easy to see whether the droplets are present in the simulation by calculating the histogram of the local field variable. For this end a hysteresis cycle was made by moving $g_{B4}$ through the values $-0.01 n$, $n=3,4,\ldots,8$ and performing twice 40000 sweeps with initial configurations $\phi_n=0$ and $\phi_n=5$ at each point. This hysteresis cycle closes at the end points, at $g_{L4}=-0.03$ and $-0.08$ but the Monte-Carlo series are not convergent in between, cf. Fig. \ref{histf}. The histogram, the probability distribution of $\phi_n$, is calculated for $\phi\in[-\Phi_0,\Phi_0]$ by dividing the interval $[-\Phi_0,\Phi_0]$ into $N_h$ subinterval and by counting the number of field variable found within each sub-interval.

\begin{figure}
\includegraphics[width=5cm,angle=-90]{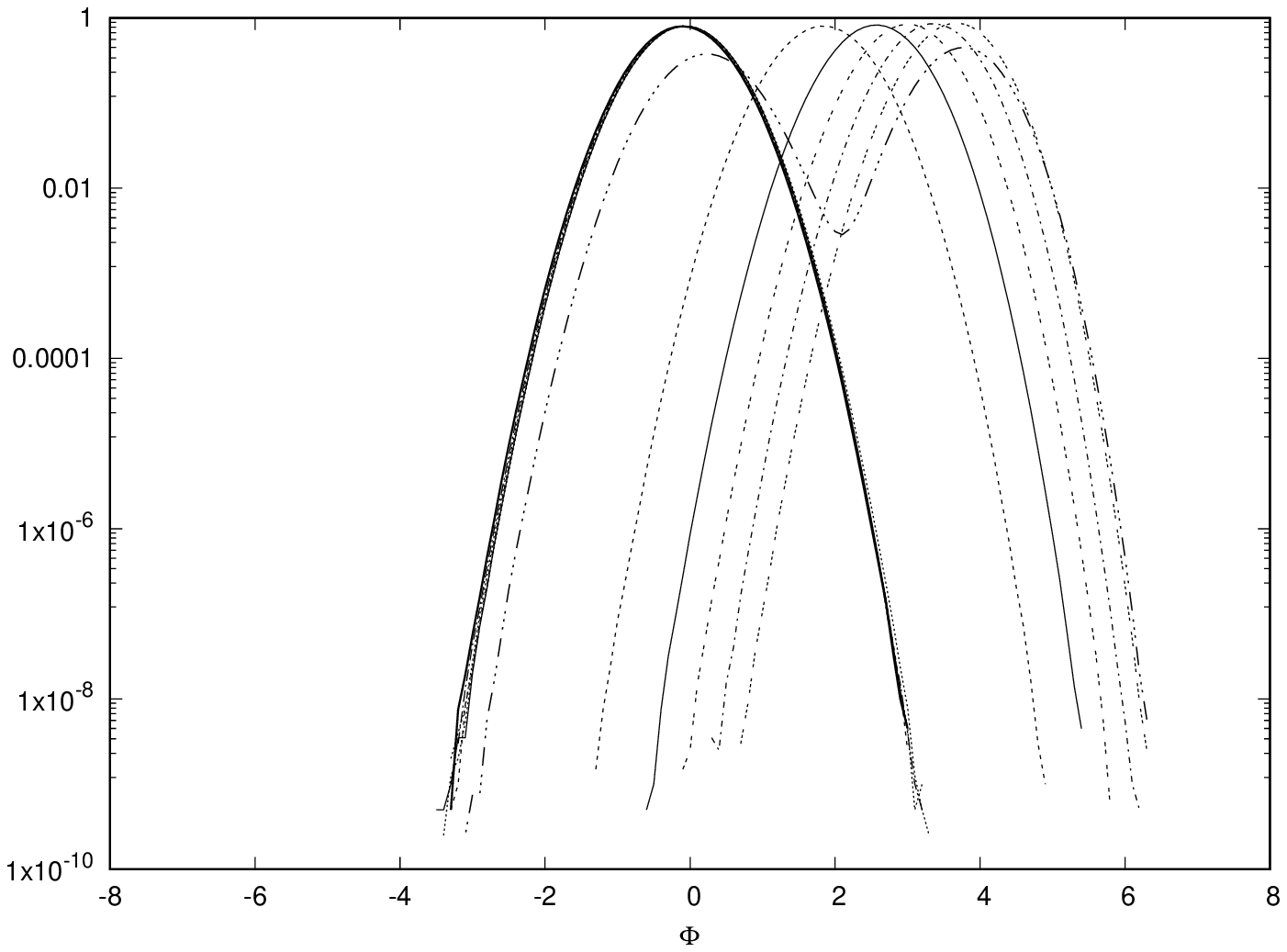}\hskip1cm
\includegraphics[width=5cm,angle=-90]{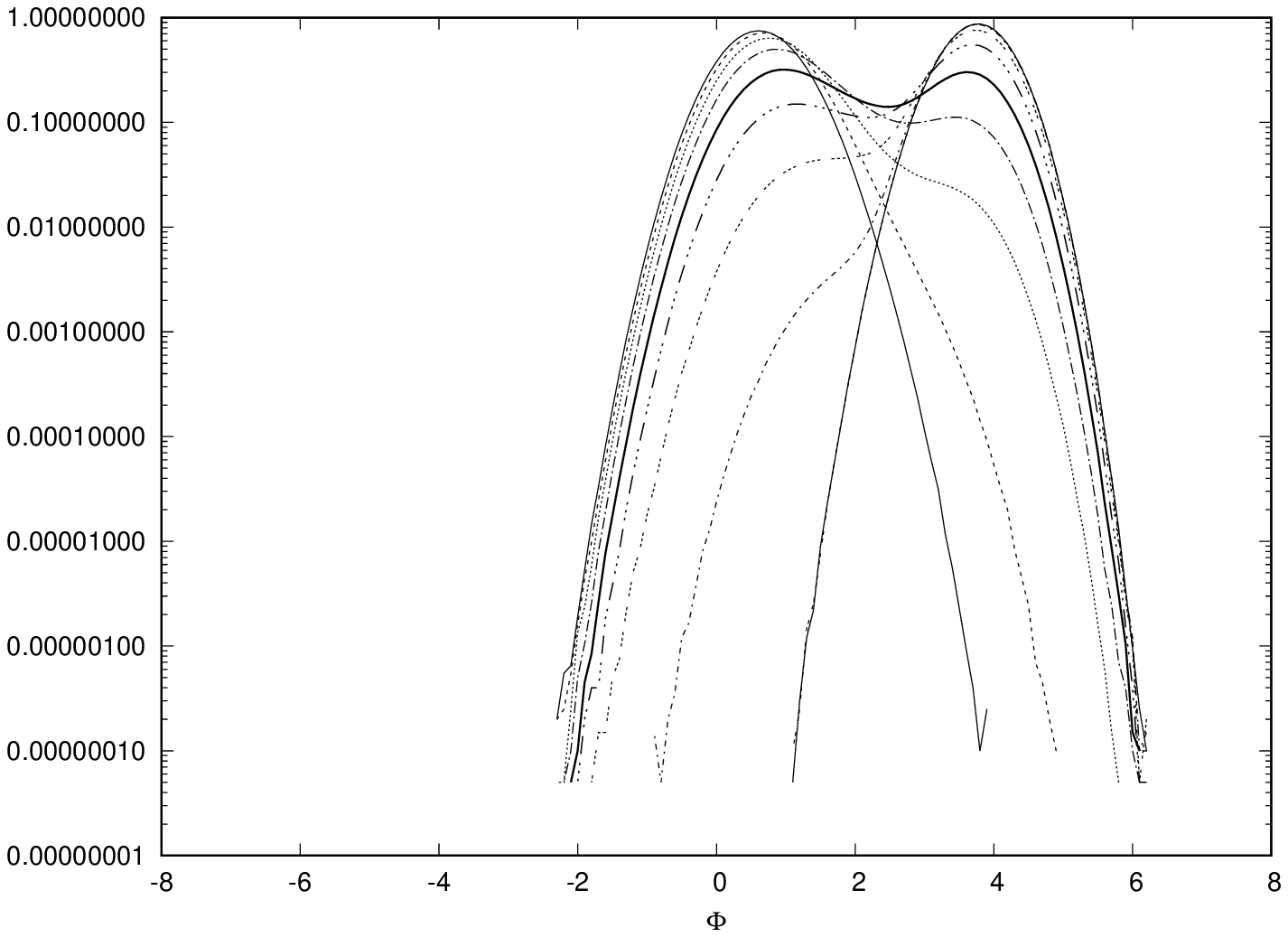}

\centerline{(a)\hskip8cm(b)}
\caption{(a): The normalized distributions of $\phi_n$. The bare parameters are $g_{L2}=0.01$, $g_{B4}=-0.01 n$, $n=3,4,\ldots,9$, $g_{L6}=0.1$. (b): The evolution of the histogram at $g_{L4}=-0.8$ with initial configuration $\phi_n=0$ by collecting statistics for the sweeps $190000+1000 n\le n_{sw}<191000+1000 n$ where $n=0,\ldots,9$. No local field was found at values without historgram curve.}\label{histogrf}
\end{figure}

The normalized histograms with $\Phi=8$ and $N_h=200$ are shown in Fig. \ref{histogrf}(a) at the six values  of $g_{L4}$. The histograms corresponding to the initial configuration $\phi_n=0$ peak around zero and are undistinguishable on the plot with linear scale for $\Phi$ except the one at $g_{L4}=-0.08$ which displays two peaks, indicating that the system found the way to the energetically stable vacuum approximatively at half-time of the iteration series. The histograms of the initial configuration $\phi_n=5$ show the gradual decrease of the average $\Phi$ as $g_{L4}$ moves in the positive direction, the last histogram at $g_{L4}=-0.03$ is already in the disordered phase and is undistinguishable from the other curves peaking around zero.

The double peak of the histogram at $g_{L}=-0.8$ with the initial condition $\phi_n=0$ shows that both ordered and disordered vacuum droplets occur in this Monte-Carlo series. To find out whether they show up in the same configuration we zoom into that part of the the Monte-Carlo iteration where $\Phi$ changes fast, the result being displayed in Fig. \ref{histogrf}(b). The part of the series where the distribution displays double peak with gradually changing maximal values proves the $\Phi$ converges by inhomogeneous droplets formation rather than by a continuous homogeneous drift of a single peak probability distribution. This suggest a qualitative similarity between the coexisting region and such subseries of the simulation where the order parameter jumps between the phases. One thus arrives at the result that the droplets are absent or play a negligible role in the long, converged Monte-Carlo series. An immediate result is the conjecture that the phase coexisting region collapses to a point in the parameter space $g_{Ln}$.

\subsection{Improving the Monte-Carlo algorithms}
No droplet was found in the relaxed Monte-Carlo series hence one expects that their inclusion should speed up the algorithm around the phase transition. But such an improvement opens a new question about ergodicity: The dynamical description of phase transitions is based on the separation of the short and the long characteristic time scales. The short microscopic time scales converge and the long time scales of certain collective modes diverge and generate discontinuities between the UV and the IR parameters in the thermodynamical limit. The phase transition is understood dynamically as an approximation where such slow modes are kept time independent. The view leads to the static characterization of of a phase transition as a particular breakdown of ergodicity and the statistical ensembles should be appropriately redefined. The troublesome point is that the slow dynamical modes are usually slow in the simulation time, too. Hence the speed up the slow modes in the simulation time may change the handling of the slow dynamical modes and lead to an inappropriate breakdown of ergodicity.

Another way to phrase the problem is to recall the difference between the bare and the renormalized parameters of the theory. On the one hand, the Monte-Carlo algorithm is defined on the level of the bare action characterizing the dynamics close to the UV cutoff. On the other hand, the slow modes to speed up are non-local and their dynamics is expressed in terms of renormalized IR parameters. Thus the proper improvement of the algorithm must include the relation between the UV and the IR parameters.

There are several tested proposals to reduce the critical slow down in the vicinity of second order phase transitions. The replacement of the Monte-Carlo iteration series by an artificial, supposedly ergodic, dynamics allows us to use different discrete the time step for different modes of the theory \cite{batrouni,duane}. The choice of larger time step for the long wavelength Fourier modes of the space-time lattice can strongly reduce the relaxation time in the vicinity of a critical point. One can introduce the update of Kadanoff's block variables and interpret the resulting random walk as a multigrid generalization of the Monte-Carlo algorithm \cite{goodman}. The direct introduction of the nonlocal update of a cluster of parallel spins \cite{swendsen} became a generally used acceleration method close to the critical point of lattice spin models. As long as the modification of a local Monte-Carlo update is restricted to scales close to the lattice spacing, as in the multigrid Monte-Carlo and the cluster update schemes, the breakdown of ergodicity remains unchanged at a second order phase transition. However the Fourier acceleration method applied for the long range collective modes opens the question whether the artificial IR dynamics breaks ergodicity just in the desired manner.

While the multigrid and the cluster algorithms improve the convergence around critical points we believe that similar non-local updates run into a problem at a first order transition. The reason can be demonstrated by a very simple cluster update step to be offered when the bare potential supports several local minima $g_{L4}<-\sqrt{6g_{L2}g_{L6}/5}$: Within a box of size $N_b$ we reflect the local field variable across $\pm\phi_m/2$, $\phi_n\to\mr{sign}(\phi_n)\phi_m-\phi_n$. The size $N_b$ is chosen to bring the phenomenological droplet action,
\be
S_d=N^3_b[U(\min(\phi_m,0))-U(\max(\phi_m,0))]+3N_b^2\bar\phi^2,
\ee
as close to zero as possible from below to reach reasonable acceptance ratio. Such a procedure gives $N_b=[3\phi_0^2/\Delta U]+1$ where $[x]$ stands for the integer part of $x$ and $\bar\phi$ is a fixed parameter. This tentative update is mentioned here only for the sake of the argument, the realistic version needs further refinement to reach better acceptance ratio. The point is that the domain where the local field variables are changed should approach the total lattice volume close to the unrenormalized tree-level phase boundary. Hence we enter into the modification of the update at arbitrary long length scales without any information about the renormalized long range dynamics which determines the true transition point.

\section{Summary}\label{summarys}
The renormalization group method was used to sketch the phase structure of the three dimensional $\phi^6$ scalar field theory. The order parameter is obtained by integrating the Wegner-Houghton equation for the bare potential as the UV cutoff is lowered. The usual method, based on the beta functions at vanishing order parameter, produces an unbounded potential from below at a finite value of the cutoff and prevents us to reach the IR regime. An improved implementation of the Wegner-Houghton equation is used which is based on a weighted average of the beta functions at the different minima of the potential and provides a stable renormalized trajectory. The absolute minimum plays an increasing role in the weighted average as the cutoff is lowered and the location of the absolute minimum of the potential at the IR end point can be identified with the expectation value of the order parameter.

One finds the Wilson-Fisher fixed point at a different location and several phase transitions by the help of this procedure. The absolute minimum of the potential stays at zero or remains at a non-zero value along the renormalized trajectories deeply in the symmetric and symmetry broken side of the first order transition, respectively. But close to the first order transition line a new radiative corrections generated first phase transitions arises. A remarkable feature of the strip between the two transitions is that the potential is locked into a degenerate minimum shape within a finite scale window of the running cutoff. The degenerate minima of the potential make this region reminiscent of usual the phase coexisting region.

The mass of elementary excitations is determined in the symmetry broken phase mainly by the order parameter and such a dynamically generated mass was found to be larger than the UV cutoff. By assuming that this result survives the Wick rotation back to real time where the UV-IR crossover marks the separation of the relativistic and non-relativistic domain one is left with the conjecture that the first order symmetry breaking is an intrinsically non-relativistic phenomenon in this model.

The symmetric phase has a surprising feature, as well, it supports a new second order transition at the spontaneous breakdown of the symmetry $\phi\to-\phi$ due to the local minima of the potential appearing for sufficiently negative quartic coupling constant. Several other phase transitions are epxected due to the competition of different terms in the potential.

The phase structure is checked by Monte-Carlo simulation but the result is inconclusive owing to the presence of lattice artefacts modifying the scaling behaviour, and the slowness of the convergence of the random walk just in the region of the parameter space where the new phase transitions may appear. It is pointed out that the known methods to speed up a local update algorithm by non-local steps have to be refined close to a first order phase transition.

We are therefore left with the conjecture of a more involved phase diagram of the $\phi^6$ model and a number of problems waiting for clarification, we mention but few of them: How does the phase structure change by extending the calculation for higher order polynomials? What kind of further phase transitions occur as the result of the competition of different non-quadratic terms in the potential? How to reach the phase mixing region of the model? Is there first order phase transition with mass below the cutoff? How to improve the Monte-Carlo algorithm to go closer to the first order transition? The first order phase transition studied here was the result of a competition between the positive $\phi^2$ and the negative $\phi^4$ term of the potential and a positive $\phi^6$ term needed to keep the potential stable. Such a model is non-renormalizable in the realistic case, in four dimensions. How to find a renormalizable model for first order transitions?

\end{document}